\documentclass{bmcart}
\usepackage[utf8]{inputenc} %unicode support
\usepackage{multirow}
\usepackage{mathtools} 
\usepackage{nameref}
\usepackage{hyperref}
\usepackage{dirtytalk}
\usepackage{amssymb}
\usepackage{pifont}
\usepackage{centernot}
\usepackage{makecell}
\usepackage{float}

\def\includegraphics{}

\startlocaldefs
\endlocaldefs

\newcommand{\cmark}{\ding{51}}
\newcommand{\xmark}{\ding{55}}

\begin{document}

%%% Start of article front matter
\begin{frontmatter}

\begin{fmbox}
\dochead{Research}


\title{Towards a Better Understanding of the Characteristics of Fractal Networks}

%%%%%%%%%%%%%%%%%%%%%%%%%%%%%%%%%%%%%%%%%%%%%%
%%                                          %%
%% Enter the authors here                   %%
%%                                          %%
%% Specify information, if available,       %%
%% in the form:                             %%
%%   <key>={<id1>,<id2>}                    %%
%%   <key>=                                 %%
%% Comment or delete the keys which are     %%
%% not used. Repeat \author command as much %%
%% as required.                             %%
%%                                          %%
%%%%%%%%%%%%%%%%%%%%%%%%%%%%%%%%%%%%%%%%%%%%%%

\author[
  addressref={aff1},                   % id's of addresses, e.g. {aff1,aff2}
%   noteref={n1},                        % id's of article notes, if any
  email={polyake@math.bme.hu}   % email address
]{\inits{EZP}\fnm{Enikő} \snm{Zakar-Polyák}}
\author[
  addressref={aff1},
  email={marcell.nagy94@edu.bme.hu}
]{\inits{MN}\fnm{Marcell} \snm{Nagy}}
\author[
  addressref={aff1,aff2},
  corref={aff2},                       % id of corresponding address, if any
  email={molontay@math.bme.hu}
]{\inits{RM}\fnm{Roland} \snm{Molontay}}

%%%%%%%%%%%%%%%%%%%%%%%%%%%%%%%%%%%%%%%%%%%%%%
%%                                          %%
%% Enter the authors' addresses here        %%
%%                                          %%
%% Repeat \address commands as much as      %%
%% required.                                %%
%%                                          %%
%%%%%%%%%%%%%%%%%%%%%%%%%%%%%%%%%%%%%%%%%%%%%%

\address[id=aff1]{%                           % unique id
  \orgdiv{Department of Stochastics},             % department, if any
  \orgname{Institute of Mathematics, Budapest University of Technology and Economics},          % university, etc
  \city{Műegyetem rkp. 3., H-1111 Budapest},                              % city
  \cny{Hungary}                                  % country
}
\address[id=aff2]{%
  %\orgdiv{Department of Stochastics},             % department, if any
  \orgname{ELKH-BME Stochastics Research Group},          % university, etc
  \city{Műegyetem rkp. 3., H-1111 Budapest},                              % city
  \cny{Hungary}   
}

%%%%%%%%%%%%%%%%%%%%%%%%%%%%%%%%%%%%%%%%%%%%%%
%%                                          %%
%% Enter short notes here                   %%
%%                                          %%
%% Short notes will be after addresses      %%
%% on first page.                           %%
%%                                          %%
%%%%%%%%%%%%%%%%%%%%%%%%%%%%%%%%%%%%%%%%%%%%%%

% \begin{artnotes}
% %\note{Sample of title note}     % note to the article
% \note[id=n1]{Equal contributor} % note, connected to author
% \end{artnotes}

\end{fmbox}% comment this for two column layout

%%%%%%%%%%%%%%%%%%%%%%%%%%%%%%%%%%%%%%%%%%%%%%
%%                                          %%
%% The Abstract begins here                 %%
%%                                          %%
%% Please refer to the Instructions for     %%
%% authors on http://www.biomedcentral.com  %%
%% and include the section headings         %%
%% accordingly for your article type.       %%
%%                                          %%
%%%%%%%%%%%%%%%%%%%%%%%%%%%%%%%%%%%%%%%%%%%%%%

\begin{abstractbox}

\begin{abstract} % abstract
The fractal nature of complex networks has received a great deal of research interest in the last two decades. Similarly to geometric fractals, the fractality of networks can also be defined with the so-called box-covering method. A network is called fractal if the minimum number of boxes needed to cover the entire network follows a power-law relation with the size of the boxes. The fractality of networks has been associated with various network properties throughout the years, for example, disassortativity, repulsion between hubs, long-range-repulsive correlation, and small edge betweenness centralities. However, these assertions are usually based on tailor-made network models and on a small number of real networks, hence their ubiquity is often disputed.

Since fractal networks have been shown to have important properties, such as robustness against intentional attacks, it is in dire need to uncover the underlying mechanisms causing fractality. Hence, the main goal of this work is to get a better understanding of the origins of fractality in complex networks. To this end, we systematically review the previous results on the relationship between various network characteristics and fractality. Moreover, we perform a comprehensive analysis of these relations on five network models and a large number of real-world networks originating from six domains. We clarify which characteristics are universally present in fractal networks and which features are just artifacts or coincidences.

%\parttitle{First part title} %if any
%Text for this section.

%\parttitle{Second part title} %if any
%Text for this section.
\end{abstract}

%%%%%%%%%%%%%%%%%%%%%%%%%%%%%%%%%%%%%%%%%%%%%%
%%                                          %%
%% The keywords begin here                  %%
%%                                          %%
%% Put each keyword in separate \kwd{}.     %%
%%                                          %%
%%%%%%%%%%%%%%%%%%%%%%%%%%%%%%%%%%%%%%%%%%%%%%


% MSC classifications codes, if any
%\begin{keyword}[class=AMS]
%\kwd[Primary ]{}
%\kwd{}
%\kwd[; secondary ]{}
%\end{keyword}

\end{abstractbox}
%
%\end{fmbox}% uncomment this for twcolumn layout

\end{frontmatter}

\section*{Introduction}\label{chapter:1}

Network science has received a great deal of research interest in the past two decades since networks can efficiently model numerous real-world structures and phenomena, including the Internet, the WWW, cellular networks, and social networks~\cite{newman2018networks}. The primary goal of network science is to better understand the structure, origin, and evolution of real networks. For example, if we aim to efficiently stop or prevent a pandemic, it is important to explore the biological structure of the virus~\cite{forster2020phylogenetic}, the mechanisms underlying the disease~\cite{prasad2020targeting} and the social interactions of communities~\cite{yum2020social,karaivanov2020social}. 
%Network science plays a major role in tackling a pandemic from various aspects~\cite{herrmann2020using,gysi2021network}.

The breakthrough in network science dates back around the millennium since the rapid and large-scale development of computer science made it possible to store and efficiently analyze complex networks~\cite{molontay2021twenty}. An observation that there are properties, which are generally present in a large number of networks regardless of their origin, was also made in these years. The most important features of networks include the scale-free property~\cite{barabasi1999emergence}
%, which can be interpreted such that most of the nodes of a network have only a few neighbors, but there is a small number of so-called hub nodes too, with a large number of connections.
and the small-world property~\cite{watts1998collective}. % In these networks, the average distances between nodes grow proportionally to the logarithm of the network size, which means in practice that every node can be reached from any other in just a few steps.

The fractality of networks is another well-studied characteristic. While the notion of fractal scaling was originally introduced in geometry, it has been extended to complex networks as well~\cite{song2005self}. The book of Rosenberg~\cite{rosenberg2020fractal} and the survey of Wen and Cheong~\cite{wen2021fractal} give an extensive overview of fractal networks. Fractal scaling was verified in various real-world networks~\cite{song2005self,first_disassortativity,gallos2007review} and has been associated with numerous important properties, such as robustness against intentional attacks~\cite{SHM_origins_sw9} and accelerated flow~\cite{diffusion}. Consequently, it is in dire need to uncover the underlying mechanisms causing fractality. Several studies have been published throughout the years that focus on the exploration of the origins of fractality~\cite{SHM_origins_sw9,max_disassortative_longrange,kitsak2007betweenness,ebc_disassortativity} without a clear consensus.

In this work, we investigate which network characteristics influence the emergence of fractality in complex networks. To this end, we review the most influential studies, and we also extend the methodological approaches in the literature. Furthermore, we propose a completely different technique to gain a better understanding of the origins of fractality, namely, we utilize the tools of machine learning. To make our findings as universal as possible, all of the aforementioned analyses rely on our large collection of real-world and model-generated networks.

The investigated characteristics that have been connected to the fractality of networks are the following:
\begin{enumerate}
    \item Yook \textit{et al.}~\cite{first_disassortativity} and Song \textit{et al.}~\cite{SHM_origins_sw9} argued that fractality originates from the disassortativity of the network and the repulsion (disconnectedness) of the hubs. 
    \item Fujiki \textit{et al.}~\cite{max_disassortative_longrange} and Rybski \textit{et al.}~\cite{fluctuation_analysis} demonstrated, using different approaches, that there is a connection between long-range anti-correlation and fractality.
    \item Wei \textit{et al.}~\cite{ebc_disassortativity} demonstrated that the distribution of edge betweenness centrality ($BC$) influences fractality and even a few edges with high $BC$ can destroy the fractal structure of a network.
    \item Csányi and Szendrői~\cite{csanyi2004fractal_sw12} were the first to draw attention to the opposing relationship between fractality and small-worldness, however, among many others, they also mention that the transition between fractal and small-world is smooth and these two properties can also be present simultaneously.
    \item Finally, Kitsak \textit{et al}~\cite{kitsak2007betweenness} argued that in fractal networks there is a weaker correlation between the degree and the betweenness centrality of the nodes than in non-fractal networks. 
\end{enumerate}

In the \nameref{chapter:2} section, we first lay the foundation of our analyses by showing how fractality can be determined in networks, presenting different fractal network models, and describing our dataset, which forms the basis of the analyses. In the \nameref{chapter:3} section, we put under the microscope the aforementioned characteristics, which have been associated with fractality, one by one, and in \nameref{section:ml} section, we use machine learning algorithms to study how the composite of the network characteristics influence fractality. Finally, in the \nameref{chapter:4} section, we summarize our findings and propose further research questions.

\section*{Foundations and preliminaries}\label{chapter:2}

In this section, we introduce the concept of fractal network, we lay the foundation of our analyses including the determination of fractality and the description of the used mathematical network models, and finally, we describe our collection of network data in detail.

\subsection*{Fractality of networks}\label{section:1.2}
Similarly to the case of geometric fractals, the fractality of networks can also be defined by the so-called box-covering method, using the length of the shortest path between two nodes as the distance metric. The method can be summarized as follows~\cite{song2005self}: The nodes of the network are partitioned into boxes of size $l_B$ in such a way that any two nodes of a box are less than $l_B$ far from each other. The minimum number of boxes needed to cover the entire network with boxes of size $l_B$ is denoted by $N_B(l_B)$. A network is defined to be fractal, if the relation of $N_B(l_B)$ and $l_B$ follows a power law, i.e.:
\[
N_B(l_B) \sim l_B^{-d_B}.
\]
The $d_B$ exponent is called the box dimension or fractal dimension of the network.

Box-covering is proved to be an NP-hard problem~\cite{song2007calculate}, therefore, there is no efficient algorithm, which could find the exact solution, i.e. the optimal $N_B(l_B)$ number of boxes. However, numerous approximating methods have been proposed, for a collection and comparative analysis of box-covering algorithms, we refer to~\cite{kovacs2021comparative}. Here, we present only one of the most widely used methods, the Compact Box Burning (CBB) algorithm, which we use later for the boxing of our networks. The method works as follows~\cite{song2007calculate}:
\begin{enumerate}
    \item \label{cbb:1} Let $C$ be the set of uncovered nodes.
    \item \label{cbb:2} Randomly choose a $c\in C$ node, and remove it from $C$.
    \item \label{cbb:3} Remove every node from $C$, which is at distance at least $l_B$ from $c$.
    \item \label{cbb:4} Repeat steps \ref{cbb:2} and \ref{cbb:3} until $C$ becomes empty. At this point, the chosen $c$ nodes form a compact box, thus no other nodes could be added to this box.
    \item Repeat steps \ref{cbb:1}-\ref{cbb:4} until the whole network is covered. 
\end{enumerate}

%In order to resolve the common problem of non-generality of results about the origins of fractality, our study relies on a large dataset, consisting of both real-world and model-generated networks. Needless to say in the study of fractal networks

\subsection*{Determination of fractality}\label{section:2.1}

The identification of the fractal nature of networks is of great importance, however, it is a very challenging task, since most of the solutions rely on visual evaluations. To avoid the uncertainty of these techniques, we apply a more automated method to determine the presence of fractality in networks.

In theory, the determination of the fractality of a network can be done by testing whether the minimal number of boxes $N_B(l_B)$ -- determined by the box-covering method -- scales as a power of the box size. A statistical framework for the detection of power law behavior in empirical data was developed by Clauset \textit{et al.} \cite{testing_fractality}, however, this framework is rarely used to quantify the fractality of networks due to the special nature of the problem. First of all, due to the NP-hard nature of the box-covering method, the use of approximation algorithms is necessary which makes the results less suitable for statistical analysis. %These algorithms have an uncertainty factor, consequently, we cannot guarantee that their outcomes are always accurate, and lower-quality data often cause misleading results.
Furthermore, for smaller networks or for those with small average distances, the number of points resulting from box covering is not large enough to obtain reliable information by these statistical tests. Moreover, the presence of different properties in networks is usually not pure, especially in real networks. It is a common phenomenon that fractal scaling holds only in an $(l_{B,MIN}, l_{B,MAX})$ range of $l_B$ \cite{testing_fractality,Rozenfeld2009FractalAT}. Often for small $l_B$ values, the power law relation prevails, while for large $l_B$ values exponential relation holds. Consequently, one has to choose a range of $l_B$ values to run the statistical tests in, which is itself a challenging task and it also reduces the sample size.

Due to the aforementioned difficulties, in practice, the most common technique for detecting the fractal nature of a network is to plot the $(l_B, N_B(l_B))$ data points on a log-log plot, fit a straight line, and decide about the goodness-of-fit by the mean squared error, the coefficient of determination or by simply looking at the plots~\cite{max_disassortative_longrange,zheng2014simple_sw2}. Obviously, these methods and the conclusions drawn from their results are highly influenced by personal decisions as it is also pointed out by Kovács \textit{et al.}~\cite{kovacs2021comparative}. Furthermore, considering a large number of networks, the visual evaluation of plots becomes impracticable. Therefore, we will use a more automated way to decide about the fractality of networks.

We use a method introduced by Takemoto \cite{determination_of_fractality1}, which takes advantage of the observation that while in fractal networks the $N_B(l_B) \sim l_B^{-d_B}$ relation holds, for non-fractal networks $N_B(l_B) \sim e^{-d_e l_B}$ is true \cite{Rozenfeld2009FractalAT}. Here, we apply a modified version presented by Akiba \textit{et al.} \cite{determination_of_fractality2}. Namely, we fit both a power law and an exponential curve in the form of the mentioned relations to the normalized $(l_B, N_B(l_B)/N)$ points, where $N$ is the number of nodes in the network. The fitting is done by excluding the point corresponding to $l_B=1$ because it is usually an outlier. The fractality can be measured by the ratio of the root-mean-square errors of the two curves:
\[
R = \frac{RMSE_{\text{power law}}}{RMSE_{\text{exponential}}}.
\]

The idea of normalization of data points and the use of RMSE allows us to compare the goodness-of-fit of different networks. % the goodness-of-fit of each curve separately, not only their ratio.

\begin{figure}[h]
    \centering
    \includegraphics[width=0.45\textwidth]{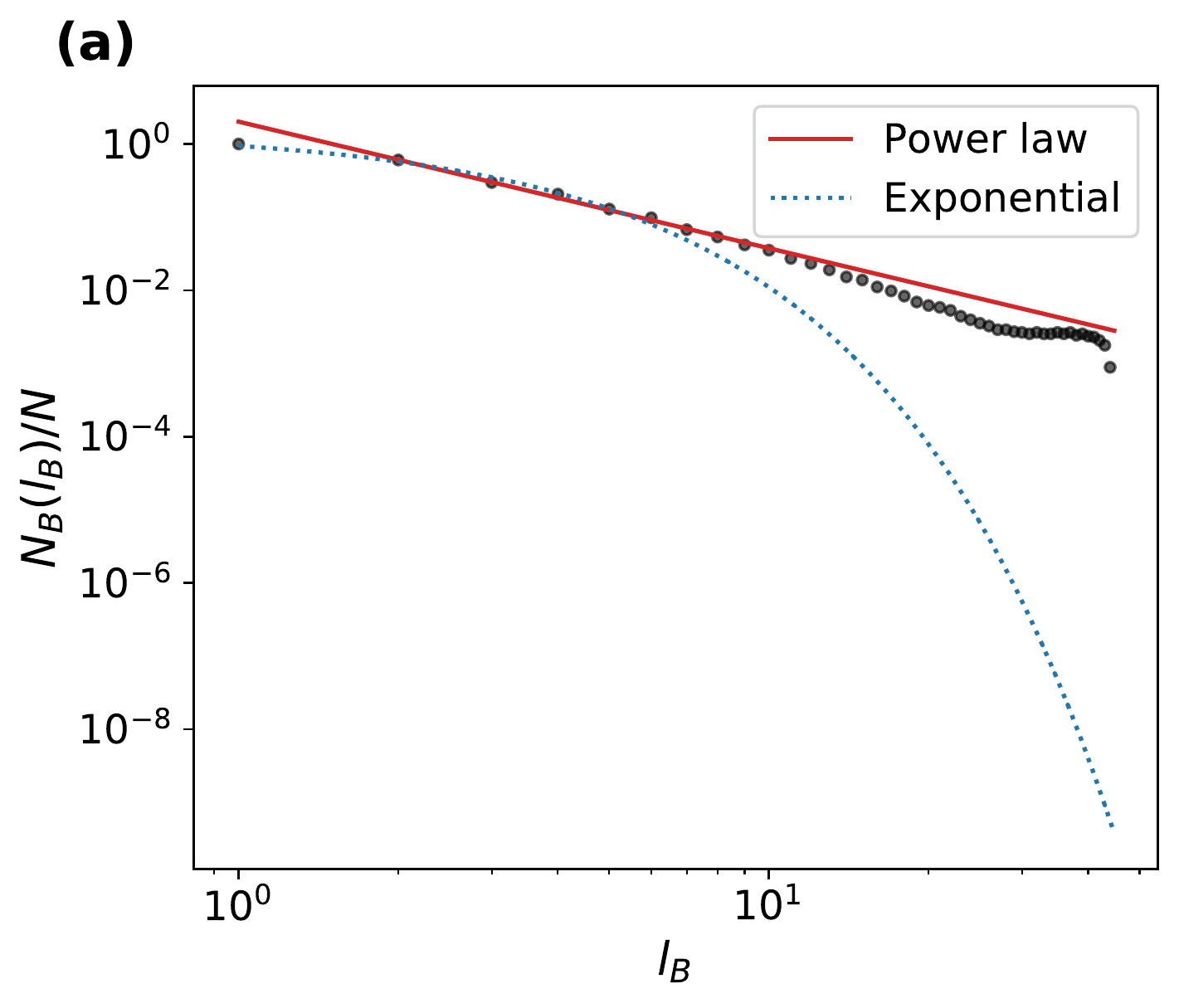}
    \includegraphics[width=0.45\textwidth]{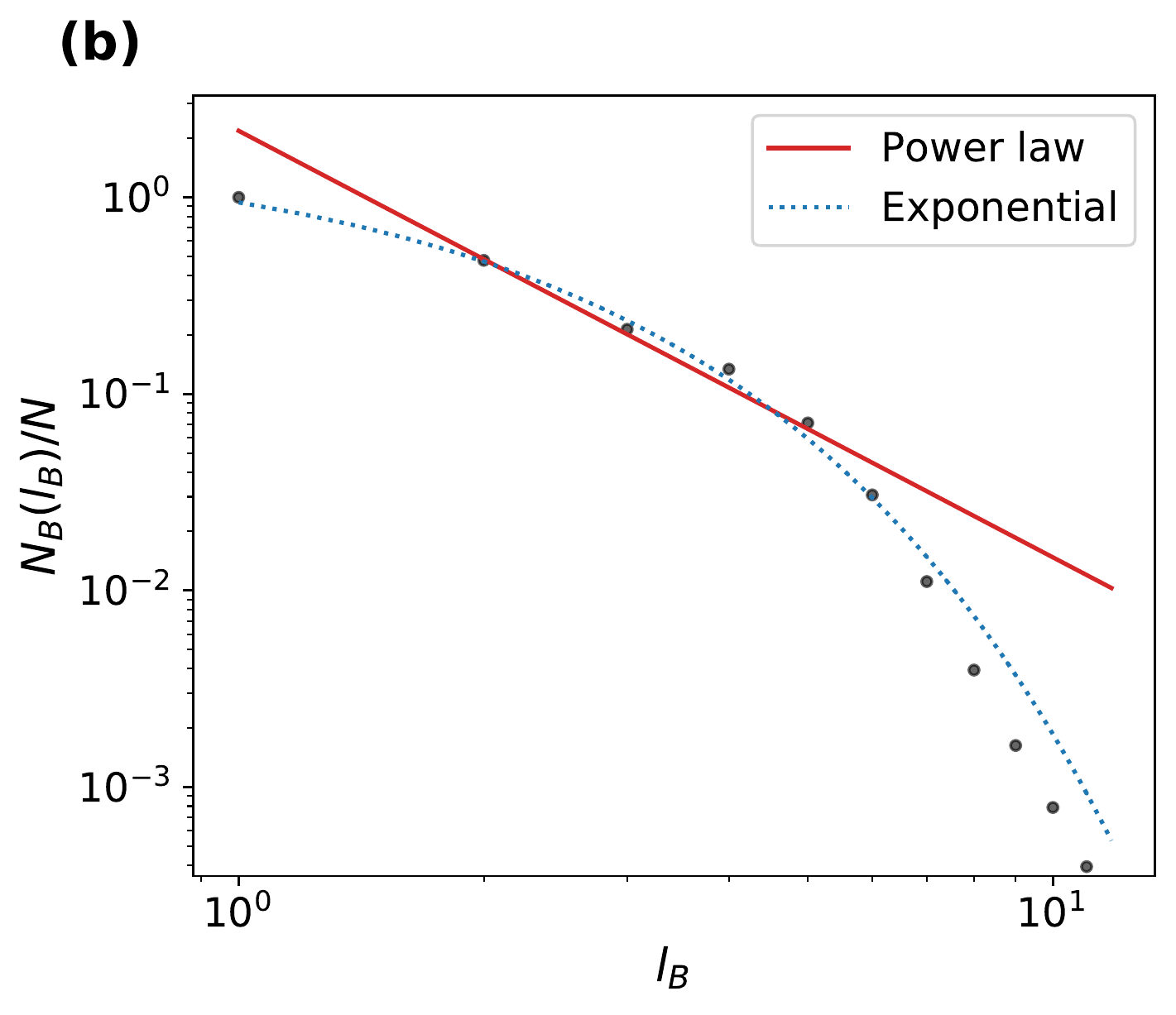}
    \includegraphics[width=0.45\textwidth]{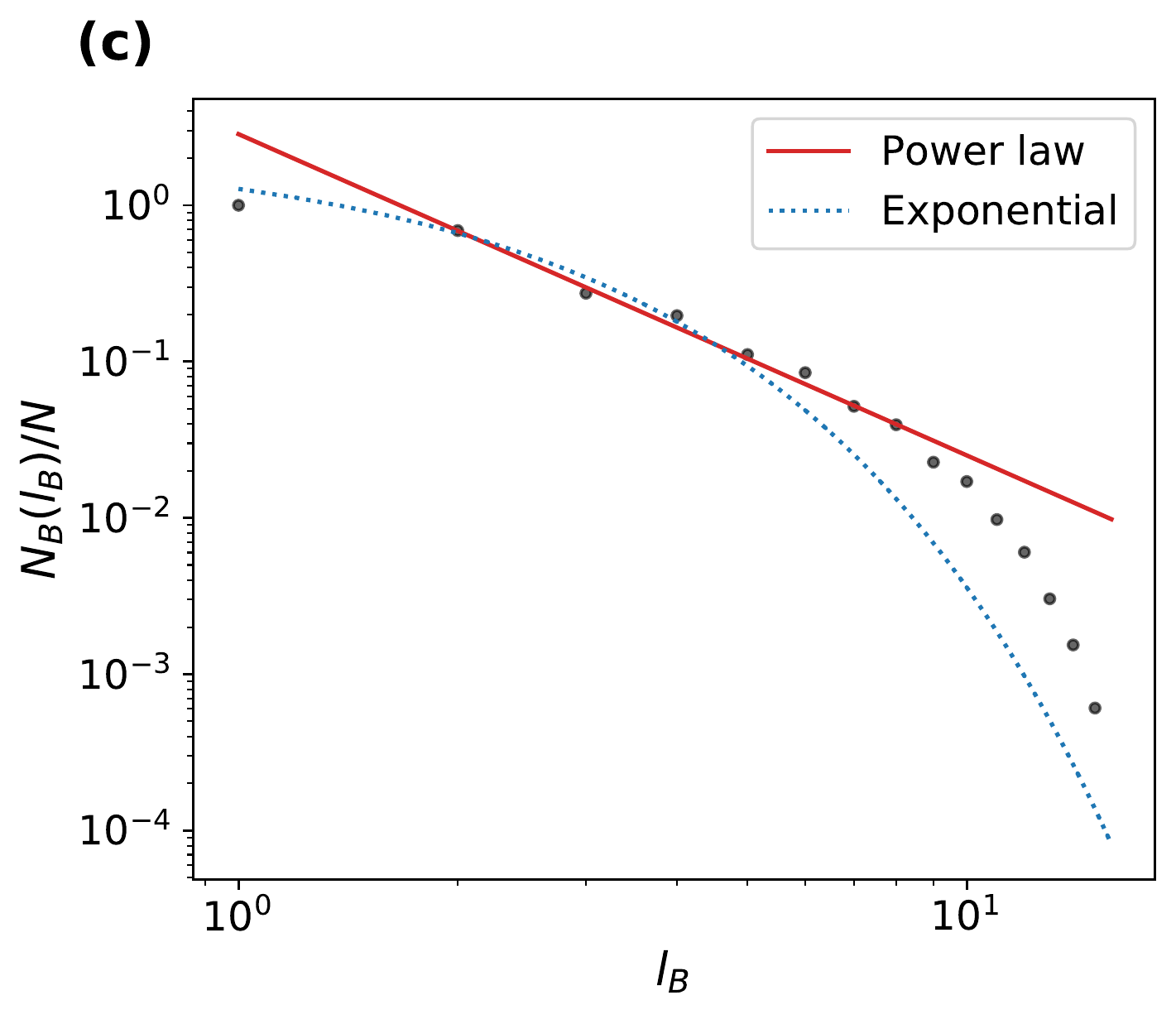}
    \caption{Results of fitting power law and exponential curves to some real networks, illustrated on log-log plots. \textbf{(a)} Brain network with $R=0.07$ \textbf{(b)} Social network with $R=3.3$ \textbf{(c)} Metabolic network with $R=0.9$}
    \label{fig:fractality_real}
\end{figure}

One might say that if $R<1$, then the network is fractal since in this case, the power law curve fits better than the exponential one, otherwise it is non-fractal. However, as was also mentioned earlier, a network is not necessarily purely fractal, but it can still possess the fractal property for a given range. This metric also allows us to measure fractality on a continuous scale, the closer the $R$ ratio is to $0$, the more fractal the network is. However, in order to compare the characteristics of fractal and non-fractal networks, we still need to create a cut-off point. We observed in both real- and model-generated networks that $R=0.65$ is a reasonable choice. Since the boundary is fuzzy, one could make a stricter partition, but for our analyses, it would not make a significant difference. Here, we say that the investigated networks with $R<0.65$ are rather fractal than non-fractal and vice versa. Figure \ref{fig:fractality_real} shows a few illustrative examples.

It is also important to note that the described method cannot be used for networks with a small diameter (e.g. smaller than 6). However, for these networks, the fractal nature can hardly be interpreted anyway. Furthermore, this method may also not give appropriate results for some mathematical network models where the fractal scaling only asymptotically holds. For this reason, we use this method for the identification of fractality only for real networks, while we stick to the theoretical findings in the case of model-generated networks.

\subsection*{Network models}\label{section:2.2}
Mathematical models play a crucial role in understanding the properties of networks. Numerous models have been introduced throughout the years to capture fractal scaling in networks and to better understand the relation between fractality and other network characteristics. In this section, we describe five such network models, with a special emphasis on the connection of their parameters with fractality.

\subsubsection*{Song-Havlin-Makse model}
One of the most well-known fractal network models is the Song-Havlin-Makse model (SHM)~\cite{SHM_origins_sw9}. The network grows dynamically and the degree correlation (hub repulsion/attraction) of the emerging graph is driven by a predefined parameter $p$. The model is defined as follows:
\begin{enumerate}
    \item \label{shm:1} The initial graph is a simple structure, e.g. two nodes connected via a link.
    \item \label{shm:2} In the iteration step $t+1$ we connect $m$ offspring to both endpoints of every edge, i.e. an $x$ node gains $m \cdot deg_t(x)$ offspring, where $m$ is a predefined parameter and $deg_t(x)$ is the degree of node $x$ at the end of step $t$.
    \item In iteration step $t+1$ every $(x,y)$ edge is removed independently with probability $p$, where $p$ is a predefined parameter. When an edge is removed, it is replaced by a new edge between the offspring of $x$ and $y$.
\end{enumerate}

\begin{figure}[h!]
    \centering
    \includegraphics[width=0.95\textwidth]{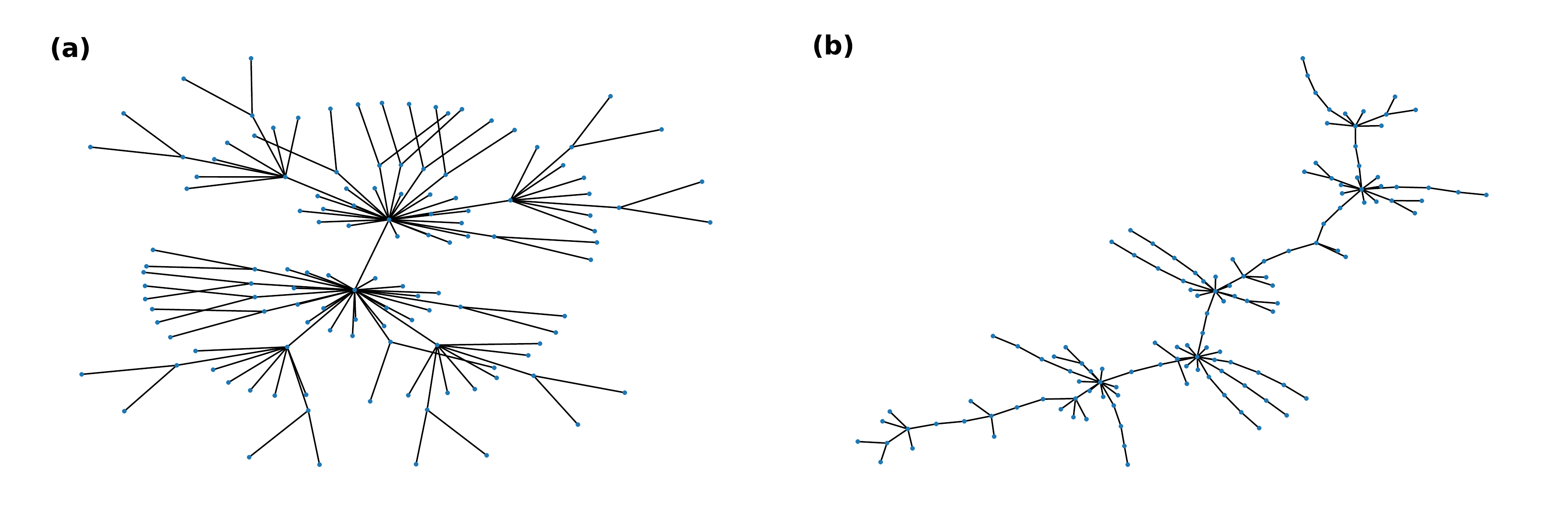}
    \caption{Illustration of the Song-Havlin-Makse model with parameter settings \textbf{(a)} $m=2, p=0$, \textbf{(b)} $m=2, p=0.6$. In both cases, three iteration steps were made.}
    \label{fig:illustration_shm}
\end{figure}

\begin{figure}[h!]
    \centering
    \includegraphics[width=0.8\textwidth]{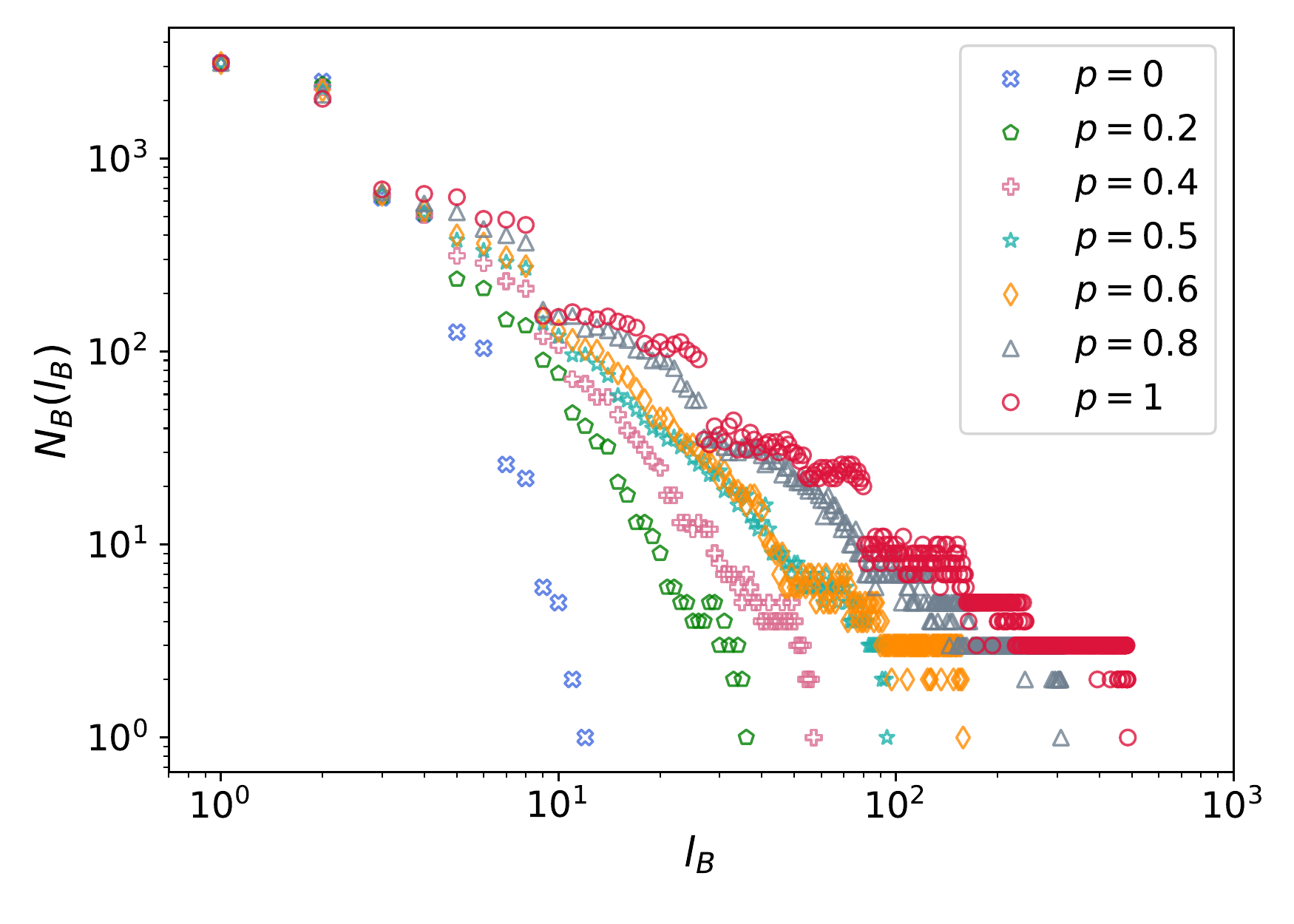}
    \caption{Illustration of the fractality of the Song-Havlin-Makse model for different parameter settings. The iteration number is set to 5, parameter $m$ is fixed to 2, and only parameter $p$ is taking different values.}
    \label{fig:boxing_shm}
\end{figure}

Figure \ref{fig:illustration_shm} illustrates two realizations that can be generated with the described model. The fractality is influenced by the choice of parameter $p$, namely, it can be shown that the generated network is fractal for $p=1$, and non-fractal for $p=0$ \cite{SHM_origins_sw9,molontay2015fractal}. The intermediate values develop mixtures between the two properties. Our observation is that networks with $p>0.6$ can be considered fractal, while those with $p<0.4$ are clearly non-fractal, which is illustrated in Figure \ref{fig:boxing_shm}. It can also be seen that the transition from fractal to non-fractal is smooth, hence in the $0.4\leq p\leq0.6$ range, it is questionable to assign the networks to any of the two categories. Later in the section called \nameref{section:ml}, where a binary classification is carried out, we are still creating a cut-off point at $p=0.5$, because we do not want to exclude the intermediate networks.

The SHM model was introduced to show that fractal networks exhibit strong repulsive relations between their hubs, which conjecture is reviewed later in the \nameref{section:assortativity_hubrep} section.

\subsubsection*{Hub attraction dynamical growth model}
Kuang \textit{et al.} modified the Song-Havlin-Makse model (SHM) in such a way that the new mechanism can generate fractal networks with strong hub attraction \cite{HADGM}. The hub attraction dynamical growth (HADG) model is based on the previously described SHM model, with the following modification applied: first, the rewiring probability of the model is flexible, more precisely, it depends on the degree of the endpoints of the links. The other modification is what they call the within-box link-growth method, which means that after an edge is rewired, the model adds additional edges between the newly added offspring, in order to increase the clustering coefficient of the network.  The evolution of the HADG model is defined as follows~\cite{HADGM}:
\begin{enumerate}
    \item The initial condition and the growth of the model are the same as in step \ref{shm:1} and \ref{shm:2} of the Song-Havlin-Makse model.
    \item \label{hadgm:2} We rewire the $(x, y)$ edge at time $t+1$ with probability % $p_{uv}$  $a$ if $\frac{deg_t(u)}{deg_t^{max}}>T$ and $\frac{deg_t(v)}{deg_t^{max}}>T$ and rewire it with probability $b$ otherwise. Formally the $p_{uv}$ edge rewiring probability of the $(u, v)$ edge at time $t+1$ is given by:
    \begin{align*}
        p_{xy}&= \begin{dcases}
        a, &\text{ if } \frac{deg_t(x)}{deg_t^{max}}>T \text{ and } \frac{deg_t(y)}{deg_t^{max}}>T\\
        b, &\text{ otherwise, }
    \end{dcases}
    \end{align*}
    where $deg_t(x)$ is the degree of node $x$ and $deg_t^{max}$ is the maximum degree in the network at time $t$ and $a, b, T \in [0, 1]$ are predefined parameters. Thus, if we define $a < b$, then hubs will have a higher probability to be connected than non-hubs.
    \item At step $t + 1$, for each old $y$ node, we add $deg_t(y)$ edges between the newly generated offspring of $y$.
\end{enumerate}
It should be mentioned that in the original paper, Kuang \textit{et al.} used the notations $a$ and $b$ for the probabilities that an edge is \textit{not} rewired~\cite{HADGM}, consequently, as a slight abuse of notation, the probabilities we use here are equivalent to $1-a$ and $1-b$ with regard to the original article.

\begin{figure}[h!]
    \centering
    \includegraphics[width=0.95\textwidth]{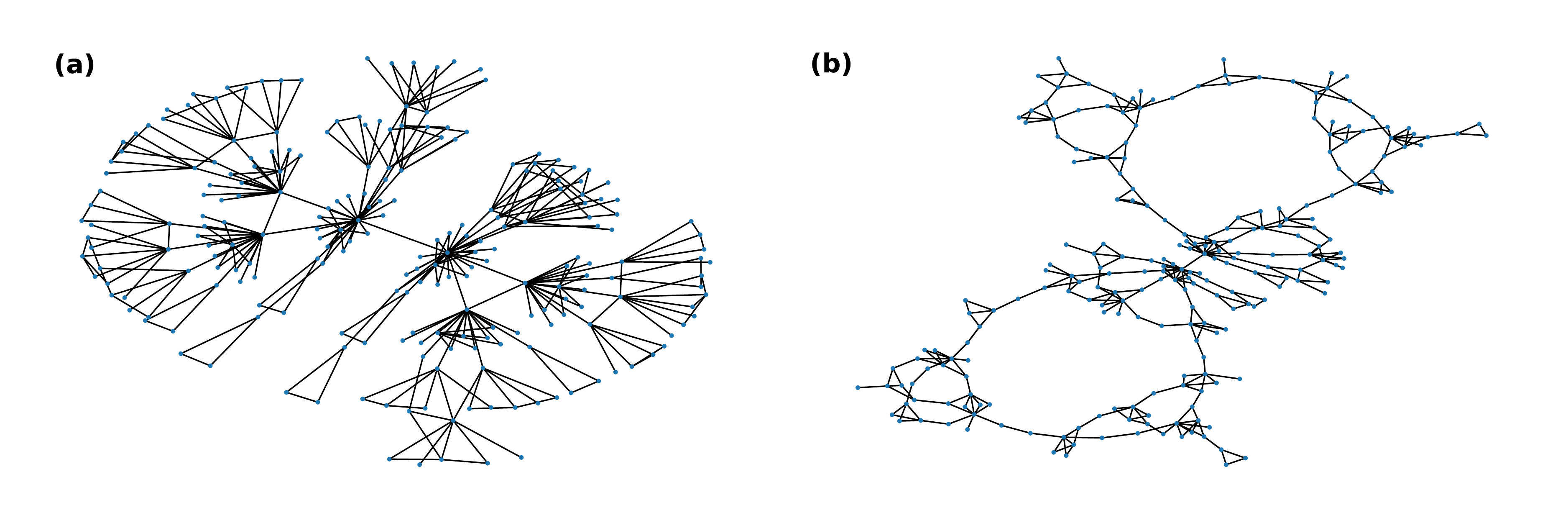}
    \caption{Illustration of the Hub attraction dynamical growth model for \textbf{(a)} $m=2, a=0, b=0.1, T=0.9$, \textbf{(b)} $m=2, a=0, b=1, T=0.9$. In both cases, three iteration steps were made.}
    \label{fig:illustration_hadgm}
\end{figure}
\begin{figure}[h!]
    \centering
    \includegraphics[width=0.8\textwidth]{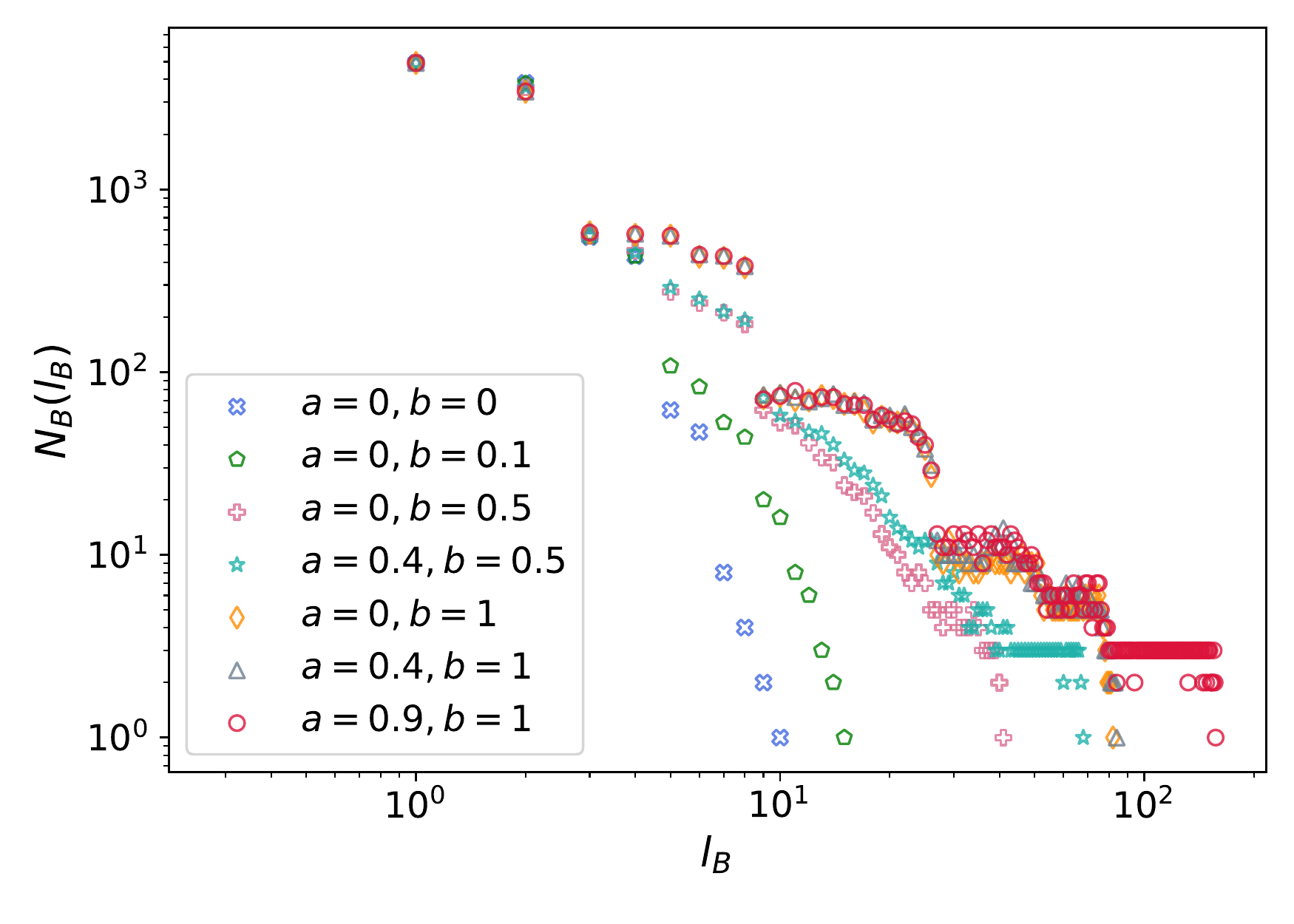}
    \caption{Illustration of the fractality of the Hub attraction dynamical growth model for different parameter settings. The iteration number is fixed to 4, parameter $m$ to 3, and $T=0.9$.}
    \label{fig:boxing_hadgm}
\end{figure}

Figure \ref{fig:illustration_hadgm} illustrates two networks that can be generated with the model using different parameter settings. Kuang \textit{et al.} concludes that there are fractal networks with assortative behavior, i.e. where the most connected nodes can be connected since this model can generate such graphs with appropriate parameter settings~\cite{HADGM}. We can support the observation of the authors, namely, we found that choosing $b>0.1$ (with our notation) results in fractal networks, and with $b\leq 0.1$ we can generate non-fractal networks, independently of parameter $a$ given that $a<b$. This is well-illustrated in Figure~\ref{fig:boxing_hadgm}.

\subsubsection*{$(u,v)$-flower}
The family of $(u,v)$-flowers was introduced by Rozenfeld, Havlin, and Ben-Avraham~\cite{uvflower_sw8}. Similarly to most of the previous models, this model also generates networks through iterations, but the edge replacement procedure is quite different. The model is defined as follows:
\begin{enumerate}
    \item The initial graph is a cycle consisting of $w=u+v$ nodes and edges, where $u$ and $v$ are predefined parameters, and we can assume that $u\leq v$.
    \item In the iteration step $t+1$ every $(x,y)$ edge is replaced by two paths connecting $x$ and $y$, one with length $u$ and one with length $v$.
\end{enumerate}

\begin{figure}[h!]
    \centering
    \includegraphics[width=0.95\textwidth]{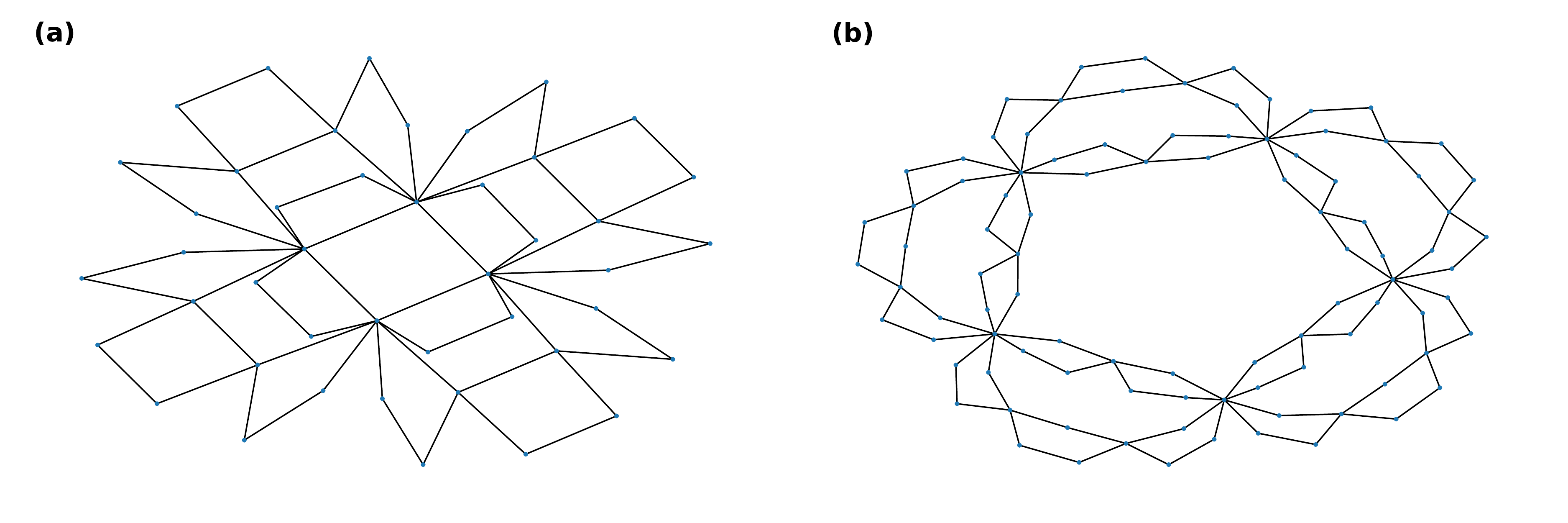}
    \caption{Illustration of the $(u,v)$-flower with three iteration steps and \textbf{(a)} $u=1, v=3$, \textbf{(b)} $u=2, v=3$.}
    \label{fig:illustration_uv}
\end{figure}

\begin{figure}[h!]
    \centering
    \includegraphics[width=0.8\textwidth]{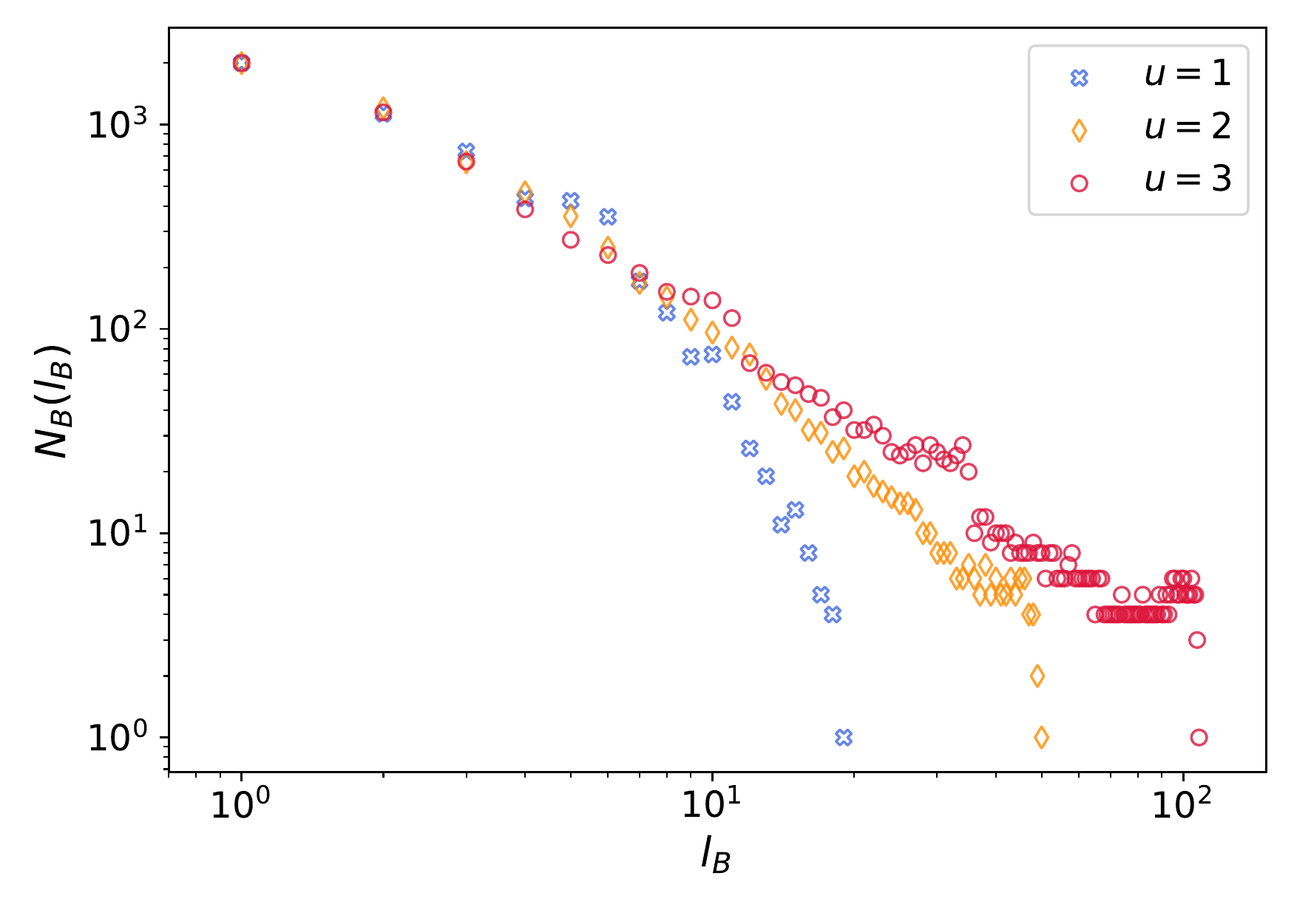}
    \caption{Illustration of the fractality of the $(u,v)$-flower for different parameter settings. In all of the cases, the iteration number is set to four, and $w=7$ is also fixed.}
    \label{fig:boxing_uv}
\end{figure}

Two networks generated with different parameter settings are shown in Figure \ref{fig:illustration_uv}. Rozenfeld~\textit{et~al}. showed that the model generates fractal networks when $u>1$, and non-fractal ones when $u=1$~\cite{uvflower_sw8}. This statement is illustrated in Figure~\ref{fig:boxing_uv}. Furthermore, it was also shown in~\cite{uvflower_sw8} that in the $u=1$ case the resulting networks are small-world, which supports the idea that fractal and small-world are conflicting properties. This statement is investigated in the \nameref{section:smallworld} section.

\subsubsection*{Repulsion based fractal model}
In \cite{zakarpolyak2022investigating} we introduced the repulsion-based fractal (RBF) model, which is also based on the SHM model~\cite{SHM_origins_sw9} and adapts some concepts of the HADG model as well~\cite{HADGM}. The model evolves through time and rewires edges with probability based on the degree of the endpoints to create repulsion among nodes. The within-box link-growth method of Kuang \textit{et al.} \cite{HADGM} is also adapted by the model to increase the clustering coefficient, hence creating more realistic networks. The growing mechanism of the repulsion-based fractal model is as follows:
\begin{enumerate}
    \item The initial condition and the growth of the model are the same as in step \ref{shm:1} and \ref{shm:2} of the Song-Havlin-Makse model.
    \item In iteration step $t+1$ we remove every edge $(x,y)$ with probability 
    \[
    p_{xy}^Y = 1 - \left|Y - \frac{\deg_t(x)+\deg_t(y)}{2\cdot \deg_{t, \max}}\right|,
    \]
    where $Y \in [0,1]$ is a predefined parameter, $\deg_t(x)$ is the degree of node $x$, $\deg_{t, \max}$ is the maximum degree at step $t$. When an edge is removed, it is replaced with a uniformly randomly chosen new edge between the offspring of its endpoints.
    \item \label{RBFM:3} We add $\deg_t(y)$ edges among the newly generated offspring of every old node $y$. In order to avoid creating self-loops, this step is only executed, when $m>1$.
\end{enumerate}

\begin{figure}[h!]
    \centering
    \includegraphics[width=0.95\textwidth]{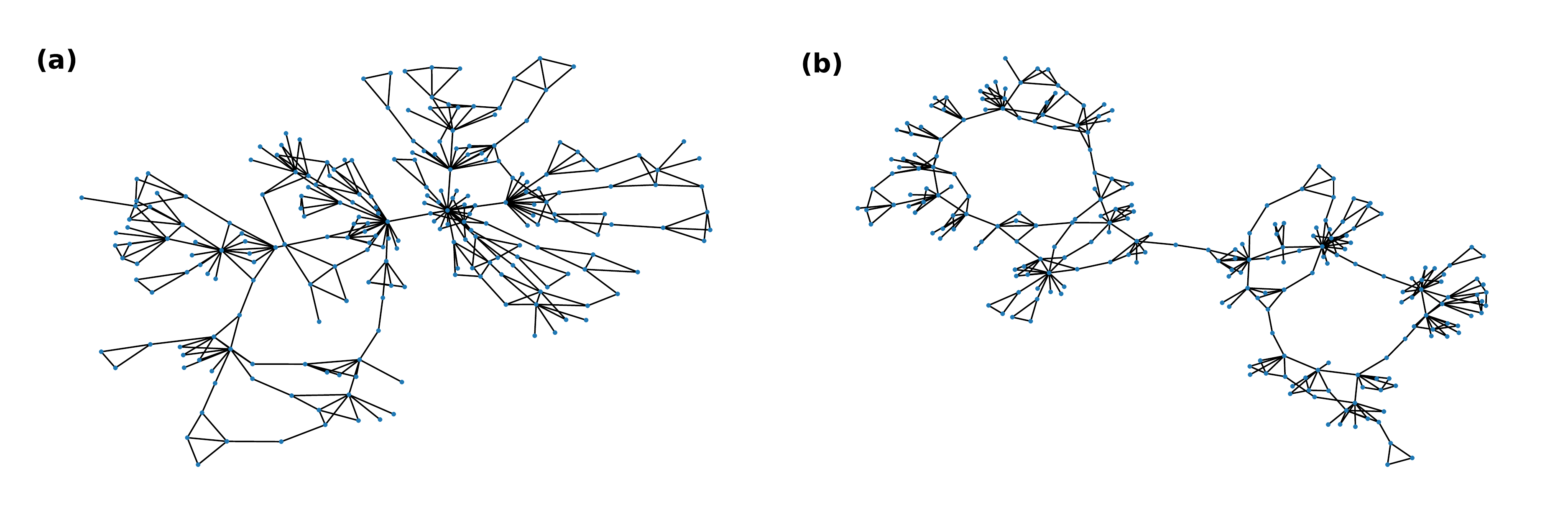}
    \caption{Illustration of the Repulsion based fractal model for \textbf{(a)} $Y=0$, \textbf{(b)} $Y=1$. In both cases, three iteration steps were made, and $m=2$ was also fixed.}
    \label{fig:illustration_rbfm}
\end{figure}

\begin{figure}[h!]
    \centering
    \includegraphics[width=0.8\textwidth]{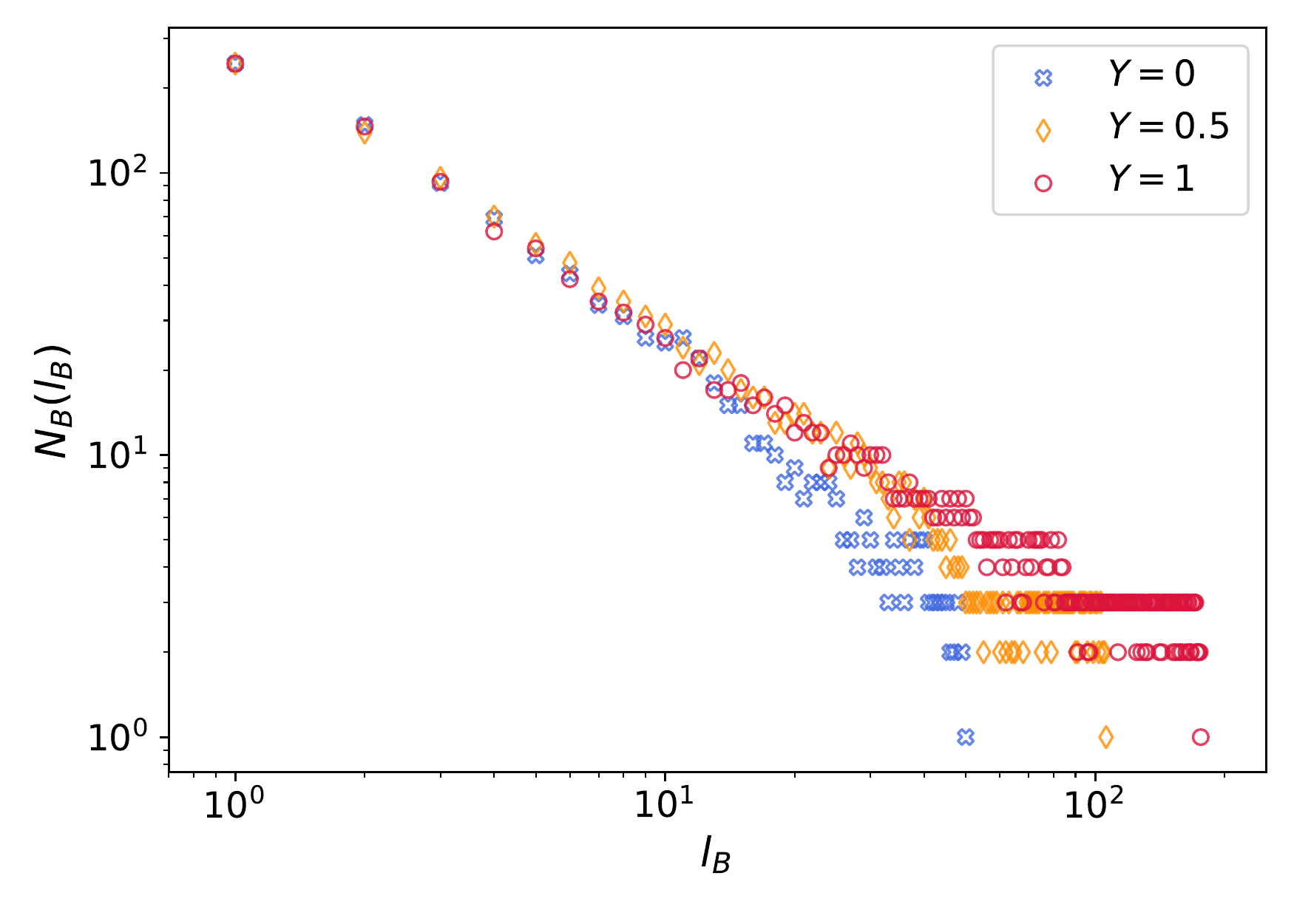}
    \caption{Illustration of the fractality of the Repulsion based fractal model for different parameter settings. The iteration number is chosen to be five, and $m=1$.}
    \label{fig:boxing_rbfm}
\end{figure}

Parameter $Y$ influences which group of nodes should repel each other (within the group). Figure~\ref{fig:illustration_rbfm} illustrates the two extreme cases of the model. This model generates fractal networks for all $Y \in [0, 1]$, as it can be seen in Figure \ref{fig:boxing_rbfm}, and hence suggests that the property, which gives rise to fractality is repulsion, but the repulsion does not necessarily have to be among hubs \cite{zakarpolyak2022investigating}.

\subsubsection*{Lattice small-world transition model}
The lattice small-world transition model (LSwTM) was also introduced in \cite{zakarpolyak2022investigating}. It utilizes the fractal nature of grid-like structures and also adapts the preferential attachment mechanism to work against fractal scaling. The model is defined as follows:
\begin{enumerate}
    \item We start with a $d$-dimensional (practically $d=2$) grid graph with $n_1 \times n_2 \times \dots \times n_k$ vertices.
    \item With probability $p$, every edge of $(x, y)$ is replaced by $(x, z)$, where $z$ is chosen with a probability that is proportional to $p_{z}$:
    \[
    p_{z} = \frac{1}{1 + \exp \left( -a \cdot \left(\frac{\deg(z)}{\deg_{\max}} - \frac{1}{2}\right)\right) }, 
    \]
    where $a$ is a positive constant, $\deg(z)$ is the degree of node $z$ and $\deg_{\max}$ is the maximum degree of the current graph.
    By default, $y$ is replaced with $z$ during the rewiring process, however, if in this way the graph becomes disconnected, $x$ is replaced instead.
\end{enumerate}

\begin{figure}[h!]
    \centering
    \includegraphics[width=0.95\textwidth]{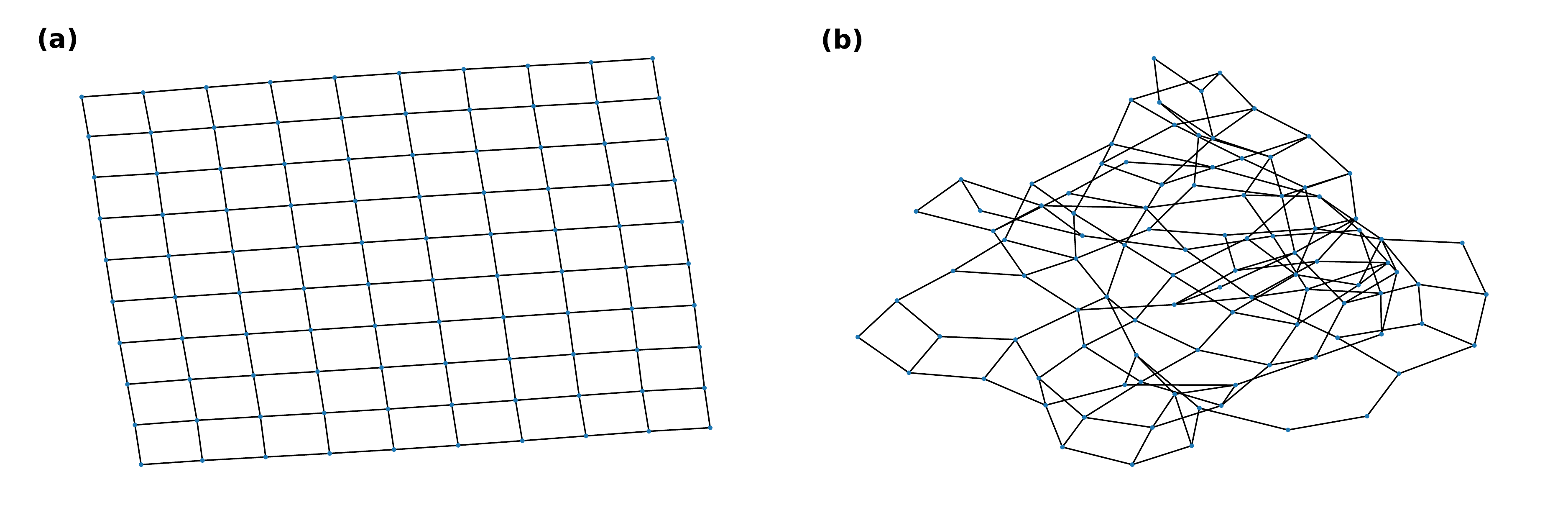}
    \caption{Illustration of the LSwTM for \textbf{(a)} $p=0$, \textbf{(b)} $p=0.1$, using a $10\times10$ lattice as the initial graph. Subfigure \textbf{(a)} is from \cite{zakarpolyak2022investigating}}
    \label{fig:illustration_mixture}
\end{figure}

\begin{figure}[h!]
    \centering
    \includegraphics[width=0.8\textwidth]{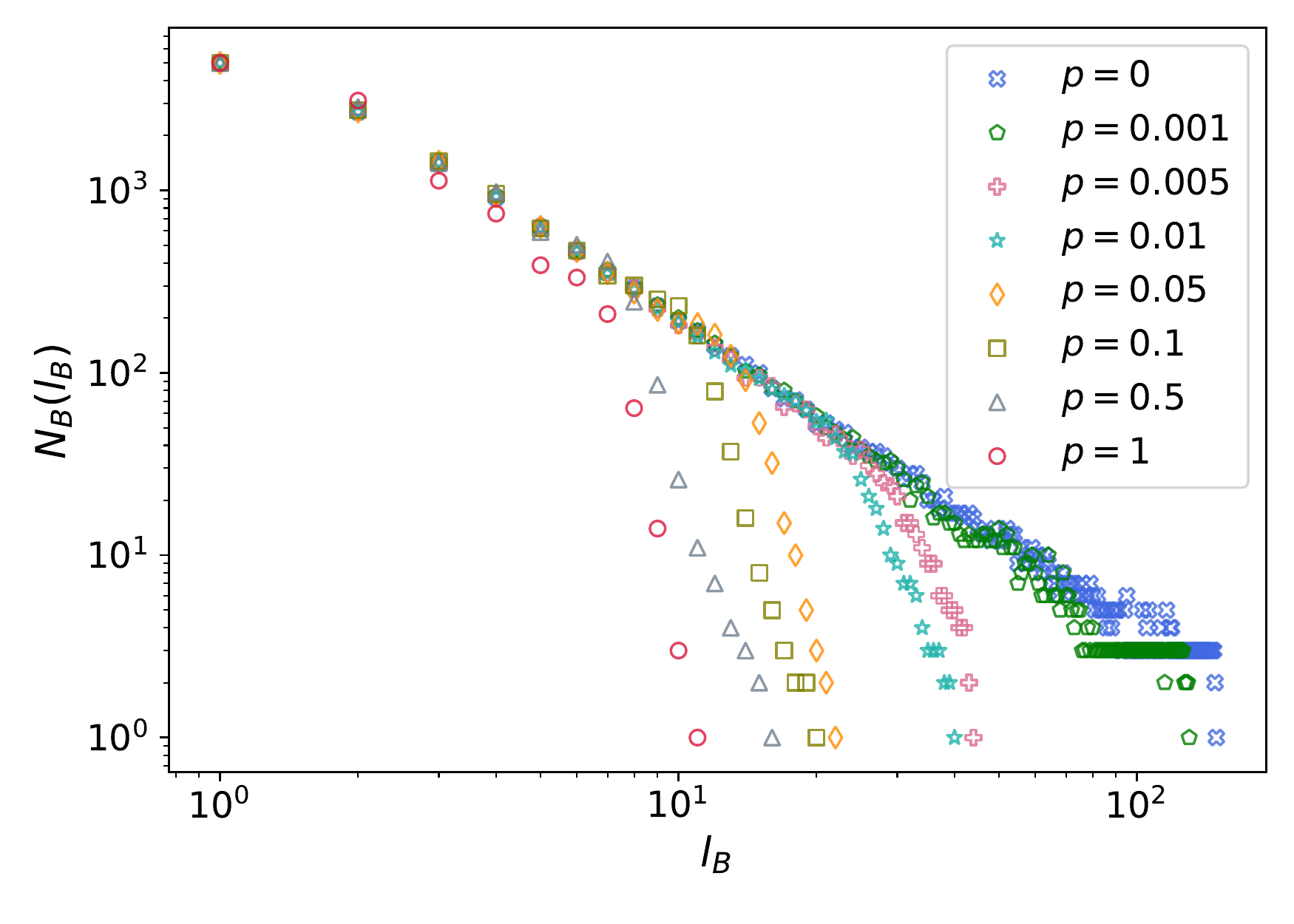}
    \caption{Illustration of the fractality of the LSwTM for different choices of parameter $p$ with $n_1=50, n_2=100$.}
    \label{fig:boxing_mixture}
\end{figure}

Even a small probability of rewiring results in a network that differs greatly from the initial grid graph, as illustrated in Figure~\ref{fig:illustration_mixture}. The fractality of the generated network depends on the choice of $p$. For $p=0$ the network is purely fractal, and as $p$ grows the model shows a transition from fractal to non-fractal networks \cite{zakarpolyak2022investigating}. As Figure \ref{fig:boxing_mixture} demonstrates, it is reasonable to choose $p=0.01$ as a cutpoint. It was also shown in \cite{zakarpolyak2022investigating} that this model demonstrates a fractal---small-world transition.

\subsection*{Data}\label{section:2.3}
To gain a complete understanding of the relationship between fractality and other network properties, it is essential to consider a diverse and large-scale collection of real-world and model-generated networks as the basis of our analysis. Although mathematical models give insight into the evolution of networks and some distinguished network properties, they usually cannot capture every characteristic of real networks. To be as comprehensive as possible, we generated networks with the models introduced in the \nameref{section:2.2} section, with various parameter settings, in addition, we collected a large number of real networks originating from six different domains. The analyses were performed in Python, and all network-related calculations, including the network generation process, were done using the NetworkX package \cite{SciPyProceedings_11}.

\subsubsection*{Model-generated networks}

We selected the parameters of the different models to get a representative sample of the space spanned by the network models while keeping the number of networks reasonably low for computational purposes. For this reason, we limited ourselves to networks with at most around 10,000 nodes.
%Fortunately, the number of nodes is deterministic in the parameters for all models, thus this limitation could be easily controlled.
%Our choices of the parameter values can be summarized as follows ($n$ denotes the number of iterations):
Our choices of the parameter values are summarized in Table~\ref{table:modelparams1}~and~\ref{table:modelparams2}.

\begin{table}[h!]
\scalebox{0.8}{
\begin{tabular}{lll||lll||lll}
\multicolumn{3}{c||}{\textbf{SHM}} & \multicolumn{3}{c||}{\textbf{HADGM}} & \multicolumn{3}{c}{\textbf{RBFM}} \\ \hline
\multicolumn{1}{l|}{$n$} & \multicolumn{2}{l||}{1, 2, 3, 4, 5} & \multicolumn{1}{l|}{$n$} & \multicolumn{2}{l||}{1, 2, 3, 4, 5} & \multicolumn{1}{l|}{$n$} & \multicolumn{2}{l}{1, 2, 3, 4, 5} \\ \hline
\multicolumn{1}{l|}{\multirow{5}{*}{$m$}} & \multicolumn{1}{l|}{1, 2, 3, 5, 10, 20, 50} & $n=1$ & \multicolumn{1}{l|}{\multirow{5}{*}{$m$}} & \multicolumn{1}{l|}{1, 2, 3, 5, 10, 20, 50} & $n=1$ & \multicolumn{1}{l|}{\multirow{5}{*}{$m$}} & \multicolumn{1}{l|}{1, 2, 3, 5, 10, 20, 50} & $n=1$ \\
\multicolumn{1}{l|}{} & \multicolumn{1}{l|}{1, 2, 3, 5, 10, 20} & $n=2$ & \multicolumn{1}{l|}{} & \multicolumn{1}{l|}{1, 2, 3, 5, 10, 20} & $n=2$ & \multicolumn{1}{l|}{} & \multicolumn{1}{l|}{1, 2, 3, 5, 10, 20} & $n=2$ \\
\multicolumn{1}{l|}{} & \multicolumn{1}{l|}{1, 2, 3, 5, 10} & $n=3$ & \multicolumn{1}{l|}{} & \multicolumn{1}{l|}{1, 2, 3, 5, 10} & $n=3$ & \multicolumn{1}{l|}{} & \multicolumn{1}{l|}{1, 2, 3, 5, 10} & $n=3$ \\
\multicolumn{1}{l|}{} & \multicolumn{1}{l|}{1, 2, 3} & $n=4$ & \multicolumn{1}{l|}{} & \multicolumn{1}{l|}{1, 2, 3} & $n=4$ & \multicolumn{1}{l|}{} & \multicolumn{1}{l|}{1, 2, 3} & $n=4$ \\
\multicolumn{1}{l|}{} & \multicolumn{1}{l|}{1, 2} & $n=5$ & \multicolumn{1}{l|}{} & \multicolumn{1}{l|}{1, 2} & $n=5$ & \multicolumn{1}{l|}{} & \multicolumn{1}{l|}{1, 2} & $n=5$ \\ \hline
\multicolumn{1}{l|}{\multirow{5}{*}{$p$}} & \multicolumn{2}{l||}{\multirow{5}{*}{0, 0.2, 0.4, 0.6, 0.8, 1}} & \multicolumn{1}{l|}{$b$} & \multicolumn{2}{l||}{0, 0.1, 0.5, 1} & \multicolumn{1}{l|}{\multirow{5}{*}{$Y$}} & \multicolumn{2}{l}{\multirow{5}{*}{0, 0.5, 1}} \\ \cline{4-6}
\multicolumn{1}{l|}{} & \multicolumn{2}{l||}{} & \multicolumn{1}{l|}{\multirow{3}{*}{$a$}} & \multicolumn{1}{l|}{0} & $b=0, 0.1$ & \multicolumn{1}{l|}{} & \multicolumn{2}{l}{} \\
\multicolumn{1}{l|}{} & \multicolumn{2}{l||}{} & \multicolumn{1}{l|}{} & \multicolumn{1}{l|}{0, 0.4} & $b=0.5$ & \multicolumn{1}{l|}{} & \multicolumn{2}{l}{} \\
\multicolumn{1}{l|}{} & \multicolumn{2}{l||}{} & \multicolumn{1}{l|}{} & \multicolumn{1}{l|}{0, 0.4, 0.9} & $b=1$ & \multicolumn{1}{l|}{} & \multicolumn{2}{l}{} \\ \cline{4-6}
\multicolumn{1}{l|}{} & \multicolumn{2}{l||}{} & \multicolumn{1}{l|}{$T$} & \multicolumn{2}{l||}{0.9} & \multicolumn{1}{l|}{} & \multicolumn{2}{l}{}
\end{tabular}}
\centering
\caption{Parameter settings for the Song-Havlin-Makse, Hub attraction dynamical growth, and Repulsion-based fractal models, with which the analyzed networks were generated. The parameter $n$ denotes the number of iterations, for the meaning of the other parameters, see the \nameref{section:2.2} section. The resulting number of networks: 138~(SHM), 161~(HADGM), 69~(RBFM).}
\label{table:modelparams1}
\end{table}

\begin{table}[h!]
\scalebox{1}{
\begin{tabular}{lll|lll}
\multicolumn{3}{c||}{\textbf{$(u,v)$-flower}} & \multicolumn{3}{c}{\textbf{LSwTM}} \\ \hline
\multicolumn{1}{l|}{$u, v$} & \multicolumn{2}{l||}{\begin{tabular}[c]{@{}l@{}}$1 \leq u \leq v$\\ $w=u+v= 2, 3, \dots, 10$\end{tabular}} & \multicolumn{1}{l|}{$N=n_1\cdot n_2$} & \multicolumn{2}{l}{\begin{tabular}[c]{@{}l@{}}10, 50, 100, 200, 500, 800,\\ 1000, 1500, 2000, 3000, 5000\end{tabular}} \\ \hline
\multicolumn{1}{l|}{\multirow{5}{*}{$n$}} & $1, 2, \dots, 9$ & \multicolumn{1}{|l||}{$w=2$} & \multicolumn{1}{l|}{\multirow{5}{*}{$p$}} & \multicolumn{2}{l}{\multirow{5}{*}{\begin{tabular}[c]{@{}l@{}}0, 0.00001, 0.0001, 0.001,\\ 0.0025, 0.005, 0.0075, 0.01,\\ 0.05, 0.1, 0.25, 0.5, 0.75, 1\end{tabular}}} \\
\multicolumn{1}{l|}{} & $1, 2, \dots, 8$ & \multicolumn{1}{|l||}{$w=3$} & \multicolumn{1}{l|}{} & \multicolumn{2}{l}{} \\
\multicolumn{1}{l|}{} & $1, 2, \dots, 6$ & \multicolumn{1}{|l||}{$w=4$} & \multicolumn{1}{l|}{} & \multicolumn{2}{l}{} \\
\multicolumn{1}{l|}{} & $1, 2, \dots, 5$ & \multicolumn{1}{|l||}{$w=5,6$} & \multicolumn{1}{l|}{} & \multicolumn{2}{l}{} \\
\multicolumn{1}{l|}{} & $1,2,3,4$ & \multicolumn{1}{|l||}{$w=7,8,9,10$} & \multicolumn{1}{l|}{} & \multicolumn{2}{l}{}
\end{tabular}}
\centering
\caption{Parameter settings for the $(u,v)$-flower and the Lattice small-world transition model, with which the analyzed networks were generated. $n$ denotes the number of iterations, for the meaning of the other parameters, see the \nameref{section:2.2} section. In the case of the LSwTM, the values of the $n_1, n_2$ parameters are chosen in such a way that $|n_1 - n_2|$ is minimal. The resulting number of networks: 118~($(u,v)$-flower), 154~(LSwTM).}
\label{table:modelparams2}
\end{table}

For those analyses, where the evaluation is done on a network-by-network basis by observing plots, we restricted ourselves to a smaller number of networks. We created three size categories of networks with approximately 800, 2000, and 5000 nodes. For every model, three to seven networks per size category were chosen including both fractal and non-fractal networks (except for the RBFM, where only fractal networks can be generated).

\subsubsection*{Real networks}

Real networks were collected from various online repositories \cite{data:biogrid, data:colorado_index, data:iwdb, data:konect, data:network_repository, data:networkscience, data:neuro, data:transportation, data:water_systems, data:wormnet, data:yeast1, data:yeast2, nagy2022network}. Table~\ref{table:data_domain} gives a short description of the different domains from which we collected the networks together with the number of networks. In total, we work with 275 real-world networks. Some of their main features are listed in Table~\ref{table:data_info}, aggregated by domains. For those analyses, which require visual evaluation, we selected four to six networks from every domain, bearing in mind to have both fractal and non-fractal networks from all size categories presented in the domain.

\begin{table}[h!]
\centering
\scalebox{1}{
\begin{tabular}{l|l|c}
\multicolumn{1}{c|}{\multirow{2}{*}{\textbf{Domain}}} & \multicolumn{1}{c|}{\multirow{2}{*}{\textbf{Description}}} & \multirow{2}{*}{\textbf{\begin{tabular}[c]{@{}c@{}}Number of \\ networks\end{tabular}}} \\
\multicolumn{1}{c|}{} & \multicolumn{1}{c|}{} &  \\ \hline
Brain & \begin{tabular}[c]{@{}l@{}}Human and animal connectomes\\ (neural connections in the brain)\end{tabular} & 45 \\
Metabolic & \begin{tabular}[c]{@{}l@{}}Protein-protein interactions of organisms\end{tabular} & 47 \\
Cheminformatics & Graph structure of enzymes & 45 \\
Infrastructural & \begin{tabular}[c]{@{}l@{}}Transportation and distribution networks\end{tabular} & 30 \\
Foodweb & What-eats-what in an ecological community & 69 \\
Social & \begin{tabular}[c]{@{}l@{}}Facebook, Twitter and collaboration networks\end{tabular} & 39
\end{tabular}}
\caption{Description of the network domains and the number of collected networks.}
\label{table:data_domain}
\end{table}

\begin{figure}[h]
    \centering
    \includegraphics[width=0.9\textwidth]{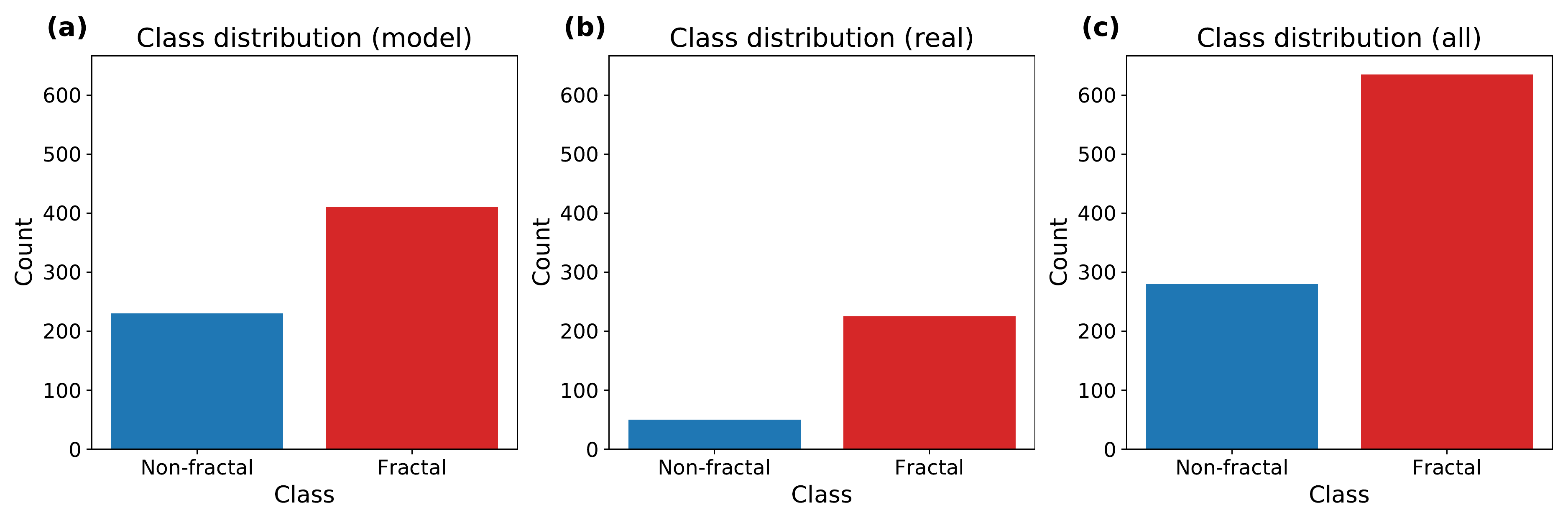}
    \caption{Class distribution concerning fractality for \textbf{(a)} model generated, \textbf{(b)} real-world, \textbf{(c)} all of the examined networks together.}
    \label{fig:class_distribution}
\end{figure}

We decided on the fractality of the networks as we described in the \nameref{section:2.1} section. In order to eliminate the randomness of the box-covering algorithm, we repeated the procedure 15 times and averaged their outcomes.
%and the network models, we repeated the procedure 15 times. More precisely, in the case of the models, for a particular parameter setting, we generated 15 networks and considered their average box counts, while for the real networks we performed the box-covering 15 times and averaged their outcomes.
The resulting class distribution of model-generated networks, real networks, and all combined networks is shown in Figure~\ref{fig:class_distribution}.
%the datasets are shown in Figure \ref{fig:class_distribution}.
It can be seen that there are much more fractal networks amongst both the model-generated and the real networks, but the number of non-fractal networks is also significant.

\begin{table}[h]
\centering
\scalebox{1}{
\begin{tabular}{l|c|c|c|c|c|c}
\multicolumn{1}{c|}{\multirow{2}{*}{\textbf{Domain}}} & \multicolumn{2}{c|}{\textbf{Number of nodes}} & \multicolumn{2}{c|}{\textbf{Number of edges}} & \multicolumn{2}{c}{\textbf{Diameter}} \\ \cline{2-7} 
\multicolumn{1}{c|}{} & \multicolumn{1}{c|}{AVG} & \multicolumn{1}{c|}{(MIN, MAX)} & \multicolumn{1}{c|}{AVG} & \multicolumn{1}{c|}{(MIN, MAX)} & \multicolumn{1}{c|}{AVG} & \multicolumn{1}{c}{(MIN, MAX)} \\ \hline
Brain & 1073 & (65, 2989) & 7485 & (730, 31548) & 14 & (2, 43) \\
Metabolic & 815 & (11, 2831) & 2099 & (10, 20448) & 9 & (1, 22) \\
Cheminformatics & 58 & (44, 125) & 100 & (77, 149) & 16 & (9, 37) \\
Infrastructural & 1268 & (8, 6474) & 2396 & (7, 15645) & 37 & (3, 122) \\
Foodweb & 115 & (19, 765) & 655 & (38, 6613) & 5 & (2, 9) \\
Social & 4465 & (86, 9763) & 7300 & (117, 24806) & 14 & (6, 28)
\end{tabular}}
\caption{Some of the main features of the collected real networks. The average, minimum, and maximum values of the number of nodes, the number of edges, and the diameter of the networks by domains.}
\label{table:data_info}
\end{table}

\section*{Analysis of network characteristics}\label{chapter:3}
In this chapter, we intend to give a comprehensive analysis of the relation of fractality with other network properties by revisiting some assertions from the literature. %When it is possible, we follow the methodology of the corresponding research papers and extend the results to a larger scale relying on our massive dataset. However, in some instances, we also extend the used methodology to investigate the problem as thoroughly as possible.

\subsection*{Disassortativity and hub repulsion}\label{section:assortativity_hubrep}
The first network properties, which were associated with the origin of fractality are disassortativity and repulsion between large degree nodes, i.e., hubs \cite{first_disassortativity, SHM_origins_sw9}. It has been much disputed whether the fractal nature of networks originates from these characteristics, there are papers that support this assertion \cite{support_disassortativity_sw6}, but there are more works that confute it \cite{ max_disassortative_longrange, ebc_disassortativity, HADGM, Masters_Marci}.

The concepts of disassortativity and hub repulsion are often used interchangeably, although the latter can be considered only as the practical interpretation of the former. For this reason, we rather separate the two notions:  First, we measure the assortativity of a network by the classic assortativity coefficient. Second, we define a novel hub connectivity score ($HCS$) as the number of edges among hubs divided by the number of hubs, thus it shows how many hub neighbors a hub has on average. Formally:
$
HCS = \frac{E_{hub}}{N_{hub}},
$
where $N_{hub}$ denote the number of hubs in the network and $E_{hub}$ is the number of edges among these hubs.
In this way, $HCS$ is large for those networks, in which hubs tend to connect to each other (strong attraction), and small, when there are only a few or no edges among them (strong repulsion).

Here, we define hubs as nodes whose degrees are at least two times the average degree of the network. If there is no such node, we set its hub connectivity score to $-1$. Furthermore, both the assortativity coefficient and the hub connectivity score are averaged over 15 realizations of the network models for each parameter setting.

\subsubsection*{Results for dissasortativity}
\begin{figure}[h!]
    \centering
    \includegraphics[width=0.95\textwidth]{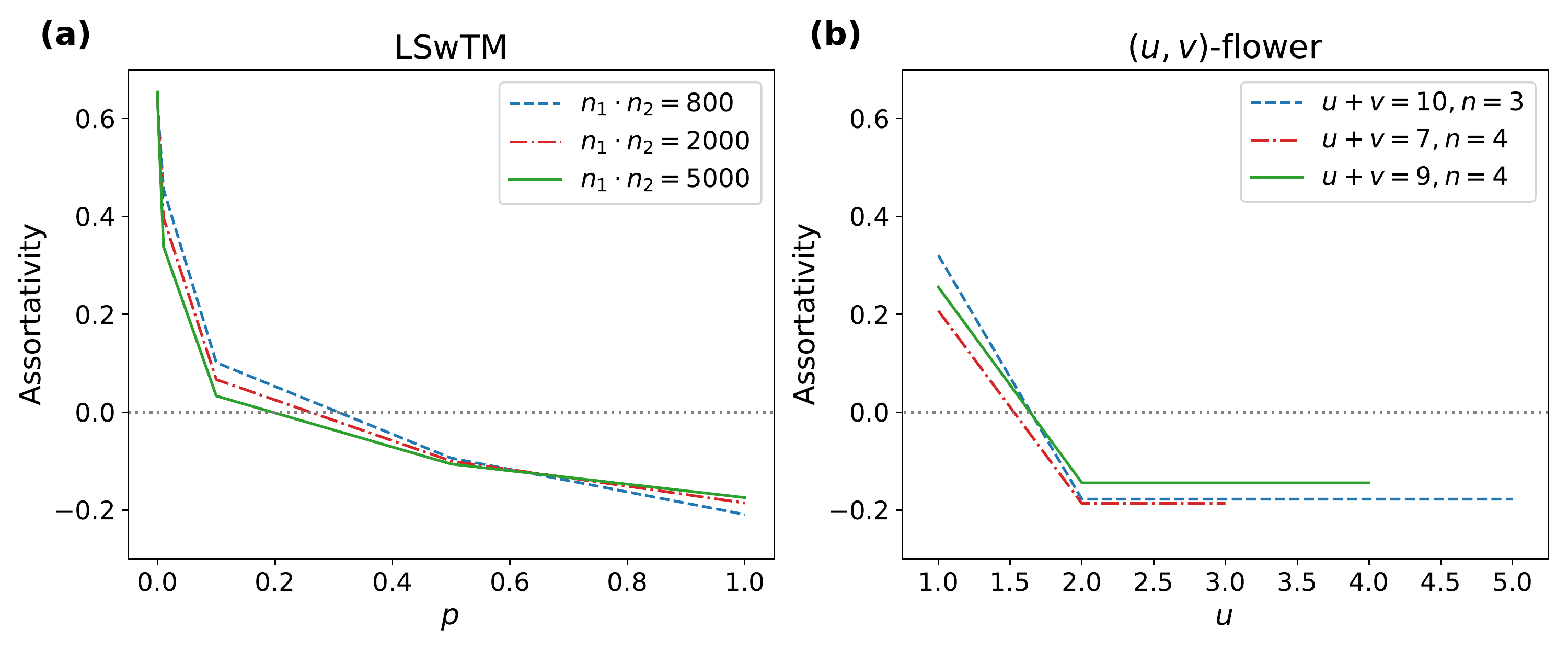}
    \caption{Assortativity as a function of the parameter, which influences fractality for the \textbf{(a)} Lattice small-world transition model, \textbf{(b)} $(u,v)$-flower. The grey dotted reference line shows Assortativity = 0. The results for the other models can be found in the supplementary material~\cite{supplementary}.} 
    \label{fig:assortativity_model}
\end{figure}

For mathematical network models, we study how the assortativity coefficient depends on the parameter of the model which influences the fractality of the network. 
%Figure \ref{fig:assortativity_model} illustrates the results on three examples for the LSwTM and the $(u,v)$-flower.
Except for the LSwT model, all models generate disassortative fractal networks, however, the $(u,v)$-flower is the only model where the fractal networks are disassortative, and the non-fractal networks are assortative, as Figure~\ref{fig:assortativity_model}~\textbf{(b)} shows. 

% With the exception of the LSwTM, all the investigated models support the statement that fractal networks are disassortative. However, it has to be mentioned that the $(u,v)$-flower is the only one that also supports that the difference in fractal and non-fractal networks can be found in their assortativity pattern: as Figure~\ref{fig:assortativity_model}~\textbf{(b)} shows, this model generates disassortative fractal and assortative non-fractal networks. 
While the fractal networks that the Song-Havlin-Makse and the Hub attraction dynamical growth models generate are disassortative, the non-fractal networks generated by these two models are also disassortative. Although in the case of the SHM model, the figures in the supplementary material~\cite{supplementary} suggest that for a specific parameter setting ($m=2$) the fractal networks are more disassortative than the non-fractal ones, for $m=1$ and $m>2$, fractal networks typically have a higher assortativity coefficient than the non-fractal ones. 
Hence, in general, based on the (dis)assortativity of the network generated by the SHM or the HADG models, no conclusions can be drawn about whether the network is fractal or not.

Similarly, the RBF model also generates disassortative fractal networks, but since the RBF model can only generate fractal networks, based on this model, no conclusions can be drawn about the assortativity of the non-fractal networks.

% While the Song-Havlin-Makse and the Hub attraction dynamical growth models also create disassortative fractal networks, no difference can be observed in the assortativity of fractal and non-fractal networks generated by these models (for illustrations see~\cite{supplementary}). 
%It has to be mentioned though, that while the $(u,v)$-flower also supports that the difference in fractal and non-fractal networks can be found in their assortativity pattern, the Song-Havlin-Makse and the Hub attraction dynamical growth model fails on this conjecture since they show that there is no connection between the fractality and the assortativity of the networks. 
The LSwTM serves as a counterexample to the aforementioned assertion because it not only generates fractal networks with assortative mixing, but we can observe a positive correlation between fractality and assortativity, i.e., the \say{more fractal} the model is, the higher the assortativity is (see Figure~\ref{fig:assortativity_model}~\textbf{(a)}). In this sense, the LSwT model behaves in the opposite way to the $(u, v)$-flower.

\begin{figure}[h!]
    \centering
    \includegraphics[width=0.95\textwidth]{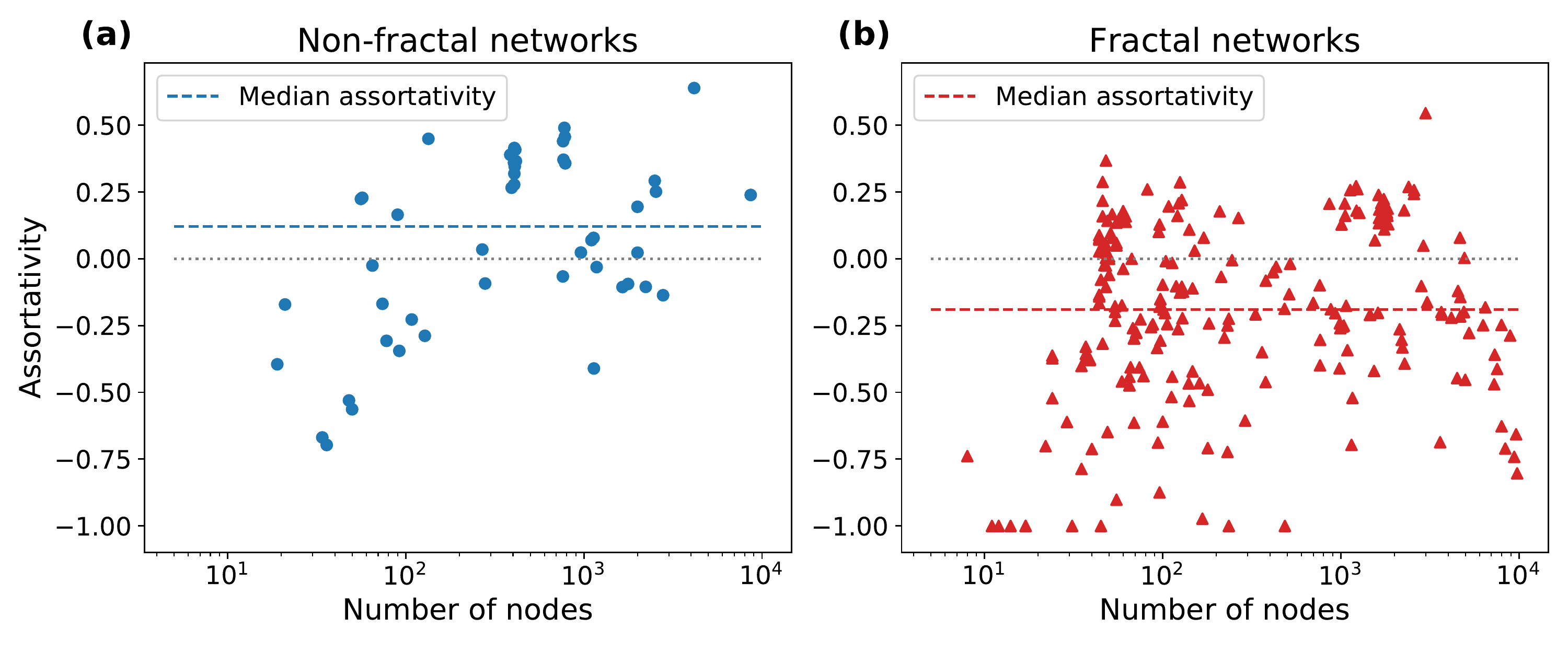}
    \caption{Assortativity of real networks plotted against the number of nodes on a semi-log scale. Subfigure \textbf{(a)} shows the non-fractal, subfigure \textbf{(b)} the fractal networks. The grey dotted reference line shows Assortativity = 0. The difference in the distribution of the assortativity of the fractal and non-fractal networks is statistically significant (Mann-Whitney U rank test's $p$-value: $2.17\cdot10^{-7}$).}
    \label{fig:assortativity_real}
\end{figure}

%Real networks do not show any remarkable pattern concerning the assortativity of fractal and non-fractal networks. Fractal networks are often disassortative, but there are numerous examples of assortative cases too, which is well illustrated in Figure \ref{fig:assortativity_real}.
In the case of real-world networks, we can say that fractal networks are often disassortative, but there are numerous examples of assortative cases too, which is well illustrated in Figure~\ref{fig:assortativity_real}. 
%The figure also shows that amongst the examined non-fractal networks there is a similar number of disassortative and assortative networks. 
Moreover, if we consider not only the binary fractal/non-fractal categories but the continuous $R$ coefficient of the networks (see the \nameref{section:2.2} section), we cannot recognize any remarkable pattern in the assortativity. An illustration of this result can be found in the supplementary material~\cite{supplementary}.

Overall, our findings partially support the conclusion of Kuang \textit{et al.}, namely, that fractality is independent of the assortative mixing~\cite{HADGM}, because there are numerous counterexamples on both sides for the conjecture that fractality originates from disassortativity. However, disassortativity is still common amongst fractal networks.
%Moreover, if we consider not only the binary fractal/non-fractal categories but the continuous $R$ coefficient of the networks (see the \nameref{section:2.2} section), we still cannot recognize any pattern in the assortativity.
%Overall, our findings support the conclusion of Kuang \textit{et al.}, namely, that fractality is independent of the assortative mixing~\cite{HADGM}. 

\subsubsection*{Results for hub repulsion}

\begin{figure}[h!]
    \centering
    \includegraphics[width=0.95\textwidth]{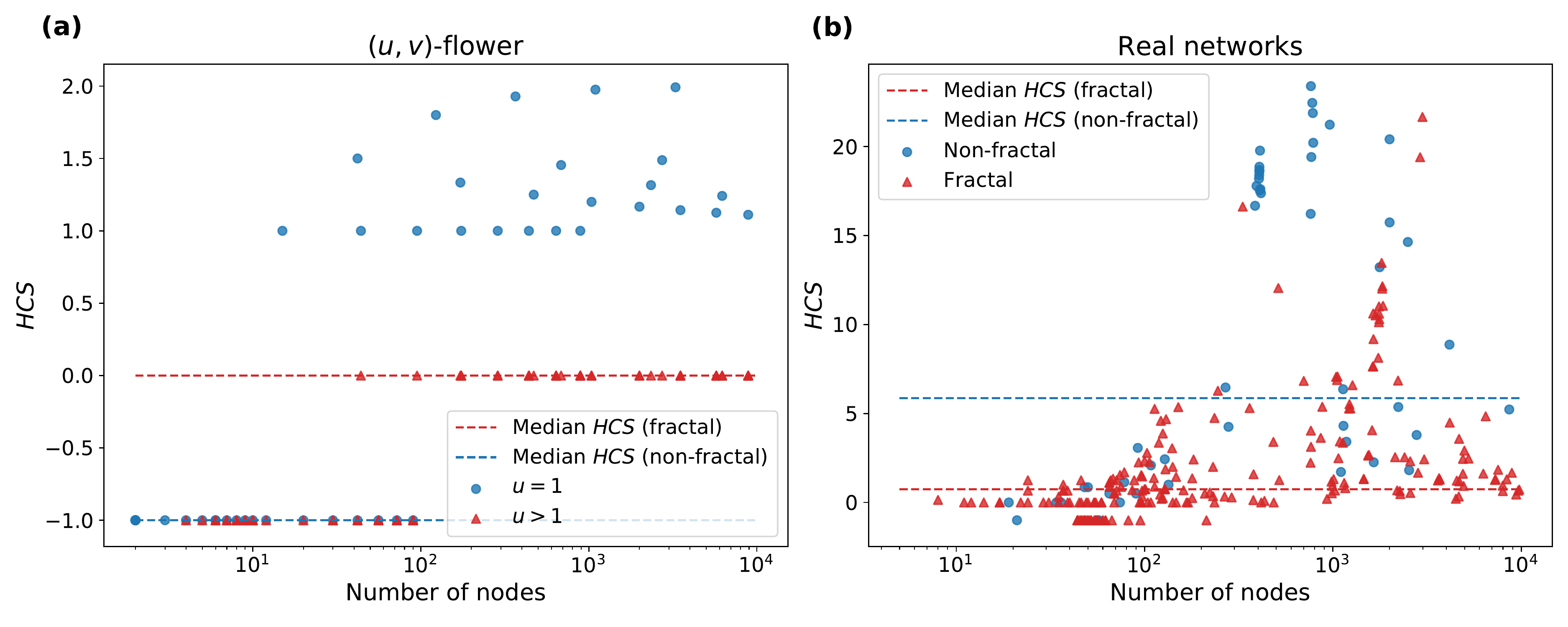}
    \caption{Plot of hub connectivity score for the \textbf{(a)} $(u,v)$-flower, \textbf{(b)} real networks. The $HCS$ is plotted as a function of the number of nodes on a semi-log scale. The difference of the distribution of the $HCS$ of the fractal and non-fractal networks is statistically significant in both cases (Mann-Whitney U rank test's $p$-value is $0.016$ for the $(u,v)$-flower, and $3.06\cdot10^{-9}$ for the real networks).}
    \label{fig:hub_rep_uv_real}
\end{figure}

Regarding the hub repulsion, we can observe that most models support the conjecture that this property may lie behind fractality. For instance, for the $(u,v)$-flower, in the $u=1$ (i.e., non-fractal) case, the HCS scores are much higher than in the fractal cases. Disregarding small networks (i.e. if the number of nodes is less than 100) due to the lack of hubs, we can say that fractal and non-fractal networks can be clearly separated according to their $HCS$. Fractal $(u,v)$-flowers have $HCS$ close to 0, while for non-fractal $(u,v)$-flowers, this measure is at least 1, which can be seen in Figure~\ref{fig:hub_rep_uv_real}~\textbf{(a)}. 

\begin{figure}[h!]
    \centering
    \includegraphics[width=0.95\textwidth]{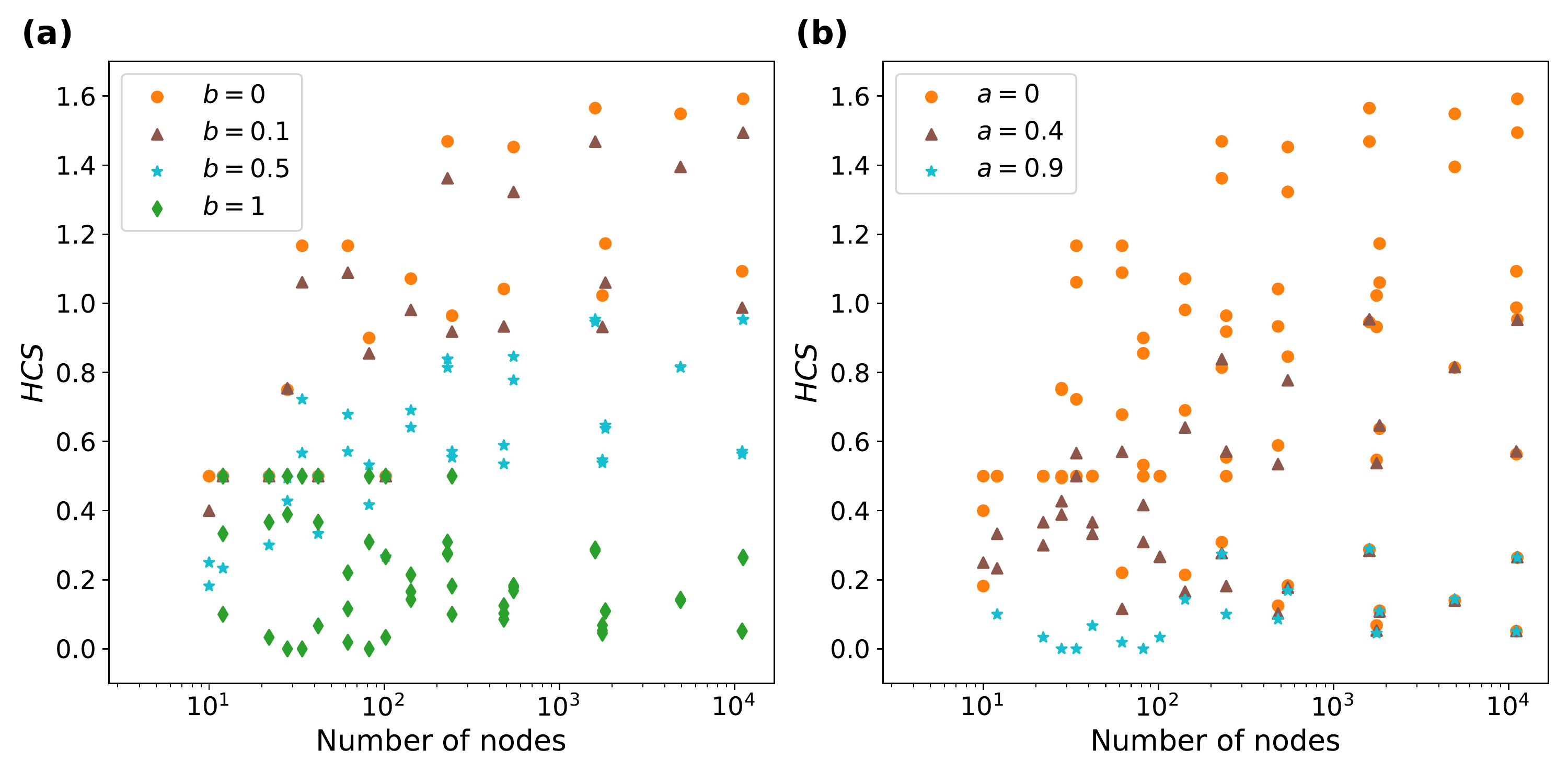}
    \caption{Plots of the hub connectivity score for the HADG model. $HCS$ is plotted against the logarithm of the number of nodes. The grouping of points is based on \textbf{(a)} parameter $b$, which influences fractality, \textbf{(b)} parameter $a$, which influences hub connectivity. }
    \label{fig:hub_rep_model}
\end{figure}

Besides the $(u,v)$-flower, the Hub attraction dynamical growth model also seems to support the conjecture. Figure~\ref{fig:hub_rep_model} shows how the hub connectivity score characterizes the different cases of the HADG model. For the non-fractal networks (i.e. when $b\leq 0.1$) the $HCS$ is higher than for the fractal networks. Furthermore, we can claim that the extent of the hub connection (or repulsion) depends on the parameter $b$, which influences the fractality of the network, and not on the parameter $a$, which creates the repulsion. Thus, as Kiang \textit{et al}.~\cite{HADGM} showed, the hubs in fractal networks can be directly connected, but our results show that the hub connectivity score is still capable of distinguishing the fractal and non-fractal networks generated by the HADG model.  

Similar behavior is demonstrated by the other models as well, that is fractal networks show stronger hub repulsion than non-fractals. However, it must be mentioned that $HCS$ usually stays between 0 and 1 for these models, and the difference concerning fractality can only be observed for each model separately. For example, the Repulsion-based fractal model is able to create fractal networks with $HCS$ being close to or even above 1, while in the case of the Song-Havlin-Makse model, only the non-fractal networks possess such high scores. Illustrations of the results for the SHM, RBF, and LSwT models can be found in the supplementary material~\cite{supplementary}.

%On the other hand, as Figure \ref{fig:hub_rep_uv_real}~\textbf{(b)} shows, real networks contradict the conjecture that hub repulsion may cause fractality. Non-fractal networks cannot be separated from fractal ones, there are as many non-fractal networks with low $HCS$ as with higher scores, and it seems that fractal networks do not necessarily have to show strong hub repulsion.
Similarly, as Figure~\ref{fig:hub_rep_uv_real}~\textbf{(b)} shows, no clear consensus can be drawn on the conjecture based on real networks. There are examples of fractal networks with large $HCS$ (i.e., strong hub attraction), 
%and non-fractal networks cannot be separated from fractal ones.
however, it can be said that networks with high $HCS$ are typically non-fractals, although there are also many examples of non-fractal networks with lower scores. Similar observations can be made if we consider the $R$ values of the networks (see: \cite{supplementary}). It is important to note that the scores are generally higher for real networks than for those that are generated by models, regardless of fractality.

In conclusion, we can say that similarly to disassortativity, strong hub repulsion is also common amongst fractal networks, but this property still cannot distinguish perfectly fractal and non-fractal networks, hence it cannot be considered as the reason behind fractality.
%In conclusion, we can say that similarly to disassortativity, hub repulsion also cannot be considered as the reason behind fractality.

\subsection*{Long-range correlation}\label{section:long_range}
Besides direct degree correlation, the long-range correlation has also been associated with fractal scaling \cite{max_disassortative_longrange, fluctuation_analysis}. Both studies suggest, based on different approaches, that there is a connection between long-range anti-correlation and fractality. Here, we apply both of the methods \cite{max_disassortative_longrange, fluctuation_analysis} in addition to a more immediate extension of neighbor-level degree correlation measures, introduced in~\cite{long_range_correlation}.

In \cite{fluctuation_analysis} a fluctuation analysis approach was proposed to measure long-range correlations. The steps of this method can be summarized as follows:
\begin{enumerate}
    \item \label{fluc_anal_1} Consider all shortest paths in the network of length $d$. For all of these paths, calculate the average degree of the nodes on the path.
    \item \label{fluc_anal_2} Calculate $F(d)$, which is the standard deviation of the mean degrees calculated in step~\ref{fluc_anal_1}.
    \item Repeat steps \ref{fluc_anal_1} and \ref{fluc_anal_2} for all possible $d$.
    \item Examine if $F(d)$ scales as a power of  $d$ with exponent $\alpha$. If so, $-\frac{1}{2}< \alpha < 0$ suggests positive, $-1 < \alpha < -\frac{1}{2}$ negative long-range correlations.
\end{enumerate}

An extension of the concept of hub repulsion to long-range scales was proposed in~\cite{max_disassortative_longrange}. The authors examined how the distribution of hub distances looks in fractal and non-fractal networks. The procedure can be summarized by the following steps:
\begin{enumerate}
    \item Calculate the distance of all pairs of hubs.
    \item For all distance $l$ calculate $\hat{P}(l)$, which is the number of hub pairs separated by the shortest path of length $l$.
    \item Calculate $\Tilde{P}(l)$ by dividing $\hat{P}(l)$ by the number of possible edges among hubs, i.e. $\Tilde{P}(l) = \hat{P}(l)/\binom{N_{hub}}{2}$.
\end{enumerate}
In this way, $\Tilde{P}(l)$ is the probability that a randomly selected pair of hubs is at distance $l$ from each other. In order to be consistent with the results of \cite{max_disassortative_longrange}, for this analysis, we cut off the hubs at the 98th percentile of the degree distribution.

%The third approach, which we consider here for capturing long-range correlations in networks was introduced in \cite{long_range_correlation}. It extends the notion of neighbor connectivity to nodes at a distance larger than one.
The third approach to capture long-range correlations was introduced in~\cite{long_range_correlation} and has not been used before to study the relationship between fractality and long-range correlation. This method extends the notion of neighbor connectivity to nodes at a distance larger than one. The main idea of the method can be summarized as follows.
\begin{enumerate}
    \item \label{mchains_1} Fix $m$, and for every node $x$, take the average degree of the nodes that are at distance $m$ from $x$.
    \item Calculate $\langle k_m\rangle (k)$ by taking the average of the outputs of step \ref{mchains_1} over nodes with degree $k$.
    \item Examine the relation of $k$ and $\langle k_m\rangle (k)$.
\end{enumerate}
Following the line of \cite{long_range_correlation}, we consider the values of $m$ up to 5, and assume power law relation between $k$ and $\langle k_m\rangle (k)$.

\subsubsection*{Results with fluctuation analysis}
\begin{figure}[h!]
    \centering
    \includegraphics[width=0.95\textwidth]{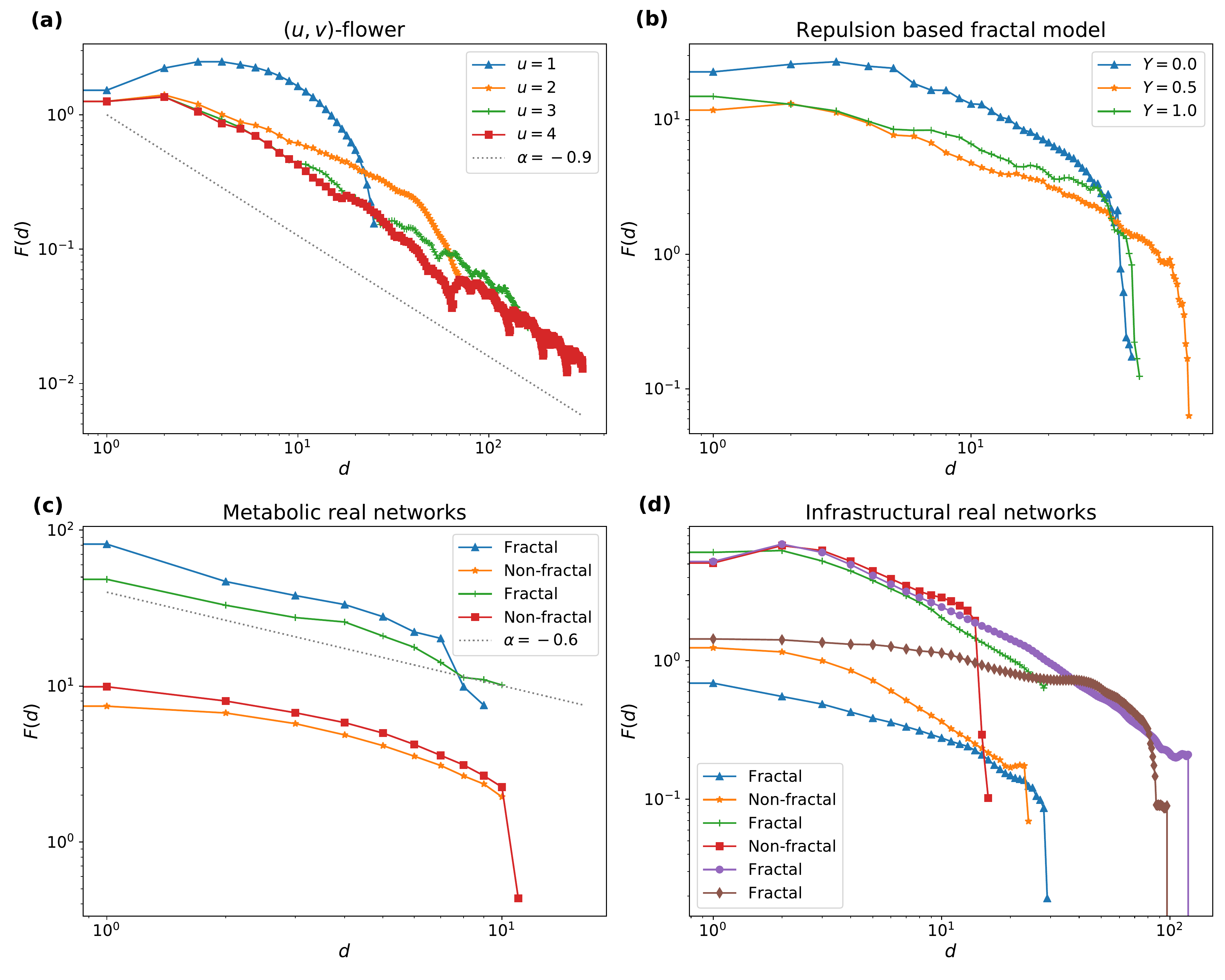}
    \caption{Results of the fluctuation analysis on a log-log scale for some examples of the \textbf{(a)} $(u,v)$-flower, \textbf{(b)} Repulsion-based fractal model. \textbf{(c)} illustrates the cases of some metabolic, \textbf{(d)} infrastructural real networks. The dotted grey lines on subfigures \textbf{(a)} and \textbf{(c)} are guides for the eye. Results for all of the examined networks can be found in the supplementary material~\cite{supplementary}.}
    \label{fig:fluctuation}
\end{figure}

The results of the fluctuation analysis are illustrated for some real-world and model-generated networks in Figure \ref{fig:fluctuation}. Generally, it can be said that for the $(u,v)$-flower and the Song-Havlin-Makse model, $F(d)$ scales as a power of $d$ with exponent less than $-\frac{1}{2}$, while in the non-fractal cases the relation is rather exponential, which supports the observations of \cite{fluctuation_analysis}. 

However, in the case of the Repulsion-based fractal model, $F(d)$ does not follow a power law. It may not be immediately visible from Figure \ref{fig:fluctuation}\textbf{(b)}, but the exponential curve provides a better fit than the power law. For the fitting, we use the \textit{powerlaw} Python package~\cite{powerlaw_package}. 

For the Hub attraction dynamical growth model, as $b$ increases, the power-law relation indeed appears, but the transition is smooth and there are fractal networks that do not show the desired relation. In the case of the Lattice small-world transition model, none of the previously mentioned relations seem to hold on $F(d)$ for the fractal networks.

Among real-world networks, there are some cases, where the expected behavior of $F(d)$ can be observed, as Figure \ref{fig:fluctuation}\textbf{(c)} shows. However, there are examples, where power law relation cannot be detected, thus long-range correlations cannot be concluded, as it is illustrated in Figure \ref{fig:fluctuation}\textbf{(d)}. In conclusion, we can say that long-range anticorrelation captured by fluctuation analysis is not a universal property of fractal networks.

\subsubsection*{Results with hub distances}
Concerning the distribution of hub distances, we can say that the HADGM, RBFM, and the $(u,v)$-flower support the suggestion of \cite{max_disassortative_longrange}, that in fractal networks hub distances have a wide range, while in non-fractal networks hubs cannot get far from each other. 

\begin{figure}[h!]
    \centering
    \includegraphics[width=0.9\textwidth]{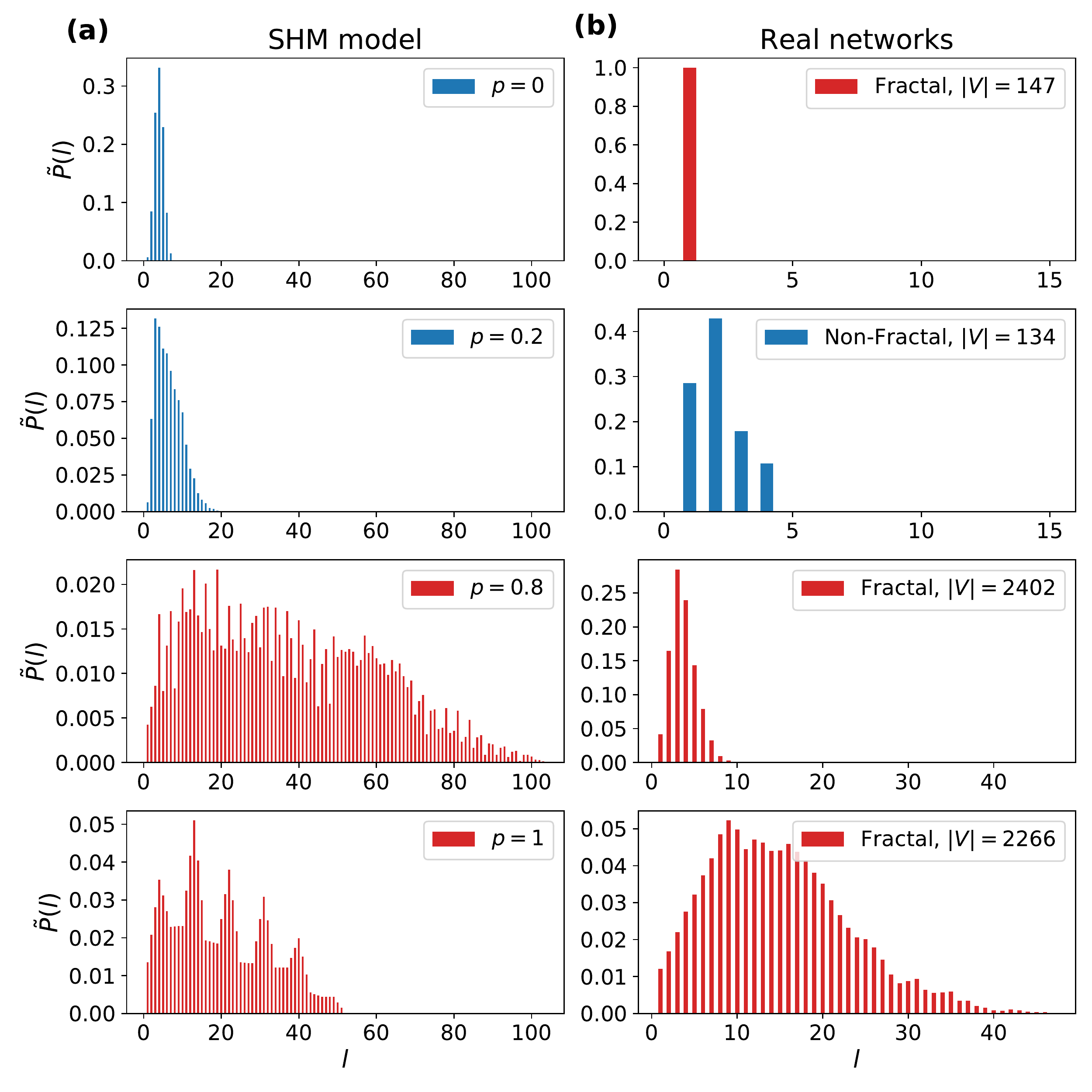}
    \caption{Distribution of hub distances for \textbf{(a)} some cases of the Song-Havlin-Makse model, \textbf{(b)} some real networks. On subfigure \textbf{(b)}, for the first two subplots, two networks with similar sizes ($\sim140$ number of nodes) are considered, and similarly to the third and fourth subplots ($\sim2300$ number of nodes). The results for the other examined networks can be found in the supplementary material \cite{supplementary}.}
    \label{fig:long_range_hub}
\end{figure}

However, a surprising observation can be made based on the Song-Havlin-Makse model. As Figure \ref{fig:long_range_hub}\textbf{(a)} illustrates, the range of hub distances expands as $p$ grows, but in the case of the pure fractal case $p=1$, it falls back to the level of intermediate networks. Consequently, this model does not seem to support that stronger fractal property always comes with larger hub distances.

In the case of the LSwTM, for $p\leq0.1$, i.e., when the model is fractal, no hubs are formed, hence this analysis cannot be carried out for this model.
%of non-fractal networks. Consequently, this model seems to contradict, that there have to be larger hub distances in fractal networks than in non-fractal ones.

Investigating real networks suggests that the examined property is independent of fractality. The first two subplots of Figure \ref{fig:long_range_hub}\textbf{(b)} show two networks of the same size where the fractal network clearly possesses smaller hub distances than the non-fractal one. %However, in most cases, both for fractal and non-fractal networks, the hub distance is small, suggesting that this property is independent of fractality.
Moreover, the third and fourth subplots of Figure \ref{fig:long_range_hub}\textbf{(b)} show two fractal networks of the same size with completely different hub distances.

\subsubsection*{Results with long-range neighbor connectivity}
Finally, the results obtained by the third approach, i.e., the neighbor connectivity~\cite{long_range_correlation}, suggest that there is no apparent connection between the fractality and the long-range correlation profile of networks. 

The Hub attraction dynamical growth model seems to be the only exception, because the non-fractal networks generated by this model usually preserve their disassortative structure for large distances as well, while fractal networks mostly do not show any correlation for distance $m\geq 3$. 

In the case of the Repulsion-based fractal model and the Song-Havlin-Makse model, usually, no correlation can be detected for $m\geq 3$ and until that, the correlation profile does not change. 

For the $(u,v)$-flower, at distance $m=3$ or $m=4$ the reverse of the correlation profile of the $m=1$ case can be observed for all networks, independently of fractality. 

\begin{figure}[h!]
    \centering
    \includegraphics[width=0.95\textwidth]{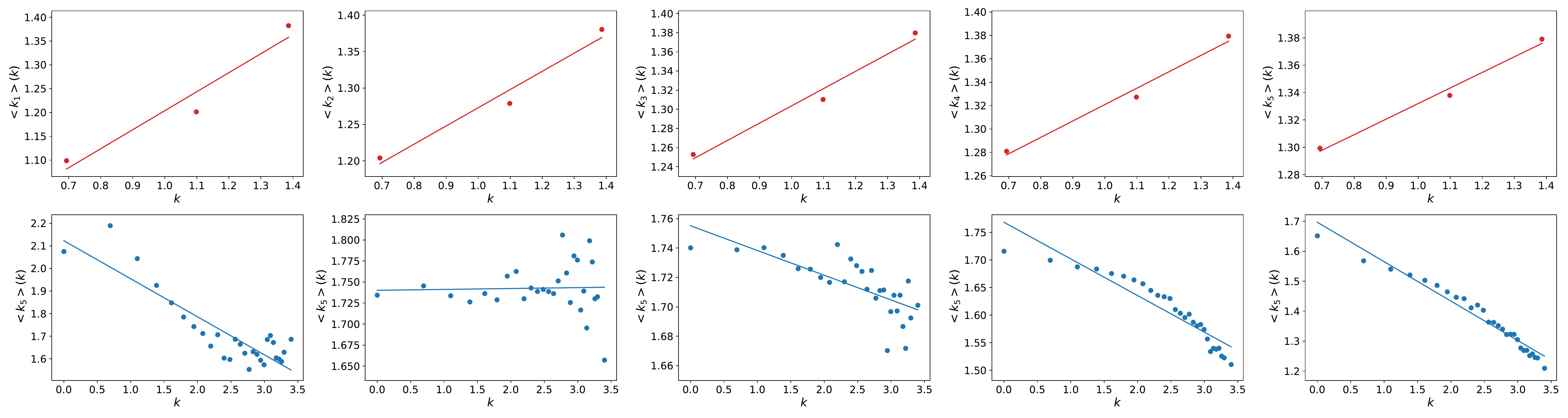}
    \caption{Degree correlations of the two extreme cases of the LSwT model at distances from 1 to 5. The first row corresponds to the $p=0$, and the second to the $p=1$ case. $\langle k_m\rangle (k)$ is plotted against $k$ on a log-log scale, and the line fitted to the log-transformed data is also provided.}
    \label{fig:mchain_mixture}
\end{figure}

Networks generated by the Lattice small-world transition model preserve their correlation profile for all $m$ distances, i.e., fractal networks have positive degree correlations, while non-fractal networks have negative correlations, even in the long-range scale (see Figure~\ref{fig:mchain_mixture}).

\begin{figure}[h!]
    \centering
    \includegraphics[width=0.95\textwidth]{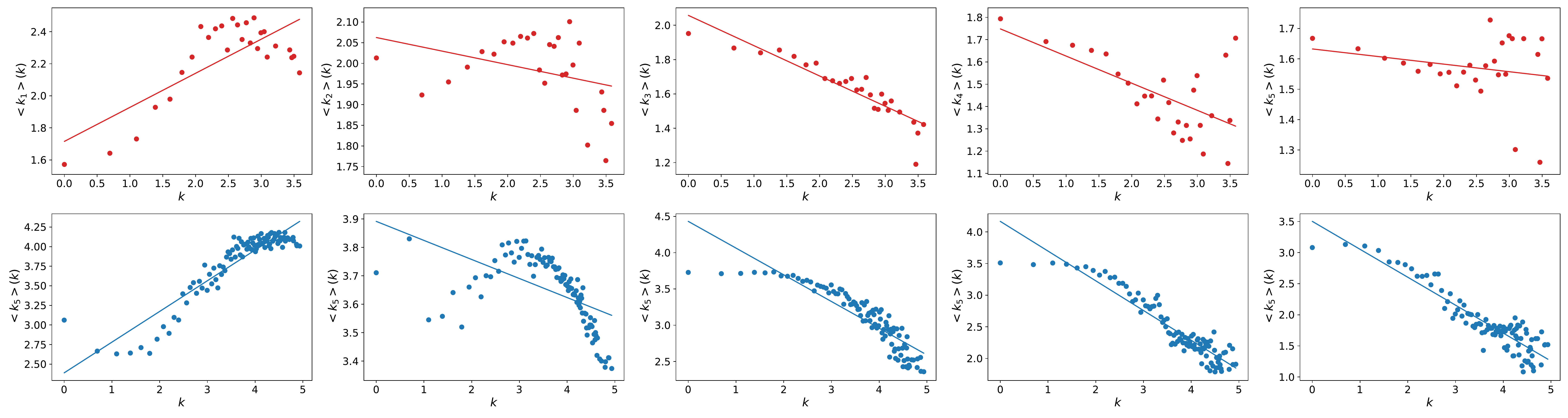}
    \caption{Degree correlations of a fractal (top row) and a non-fractal (bottom row) brain network at distances from 1 to 5. $\langle k_m\rangle (k)$ is plotted against $k$ on a log-log scale, and the line fitted to the log-transformed data is also provided.}
    \label{fig:mchain_brain}
\end{figure}
\begin{figure}[h!]
    \centering
    \includegraphics[width=0.95\textwidth]{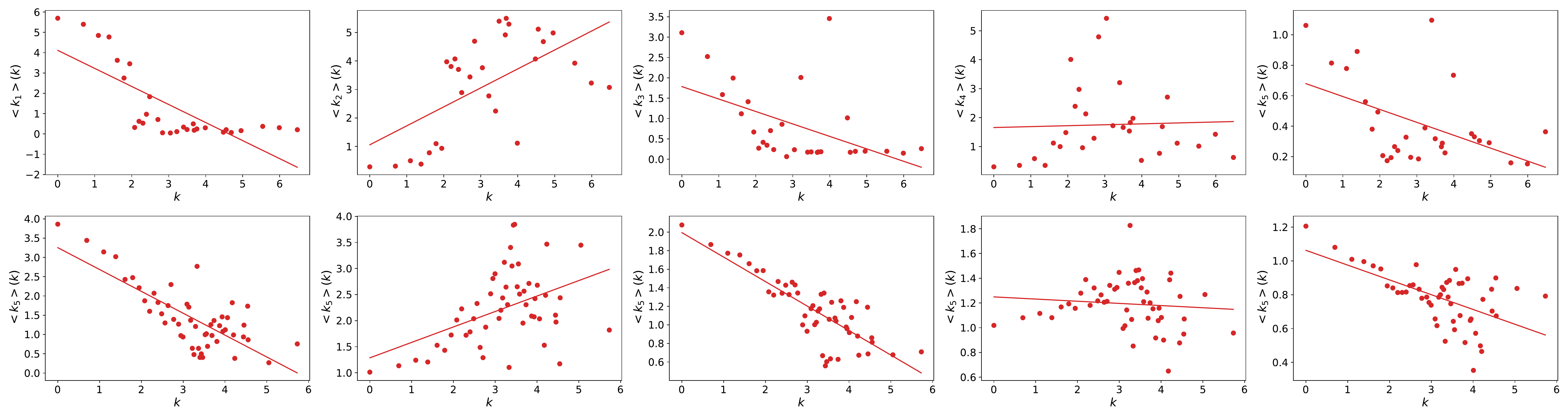}
    \caption{Degree correlations of two fractal social networks at distances from 1 to 5. $\langle k_m\rangle (k)$ is plotted against $k$ on a log-log scale, and the line fitted to the log-transformed data is also provided.}
    \label{fig:mchain_social}
\end{figure}

Degree correlations of the real networks -- independently of their fractality -- are usually preserved or reversed for larger distances but do not seem to disappear. Figure \ref{fig:mchain_brain} shows two brain networks, one of them is fractal, the other one is not, and their correlation profile is very similar for all $m$ distances. Figure \ref{fig:mchain_social} shows two fractal social networks with a negative correlation on the direct neighbor level and a positive correlation at $m=2$. Illustrations of the results for all of the examined networks can be found in the supplementary material~\cite{supplementary}.

Overall, we can conclude from the results of all three approaches that fractality and long-range correlation profiles do not have a clear ubiquitous connection.

\subsection*{Edge betweenness centrality}\label{section:ebc}
In \cite{ebc_disassortativity}, the authors reported that even a small number of edges with high betweenness centrality ($BC$) can destroy the fractal scaling of a network. Although, they investigated this conjecture from the perspective of minimum spanning trees, here we rather study the suggestion explicitly on the networks. In other words, we examine the question of whether fractal networks can have edges with high betweenness centrality. To this end, we calculate multiple measures concerning the edge betweenness centralities: the average and maximum $BC$ and the average of the top 5\% of edge betweenness centralities. We examine whether fractal networks tend to possess smaller values of the aforementioned measures.

\subsubsection*{Results}
Some of the investigated models can only generate networks with edges having a small betweenness centrality, other models can also create networks with edges with quite large $BC$ as well, therefore a comparison between models cannot be made. 

\begin{figure}[h!]
    \centering
    \includegraphics[width=0.93\textwidth]{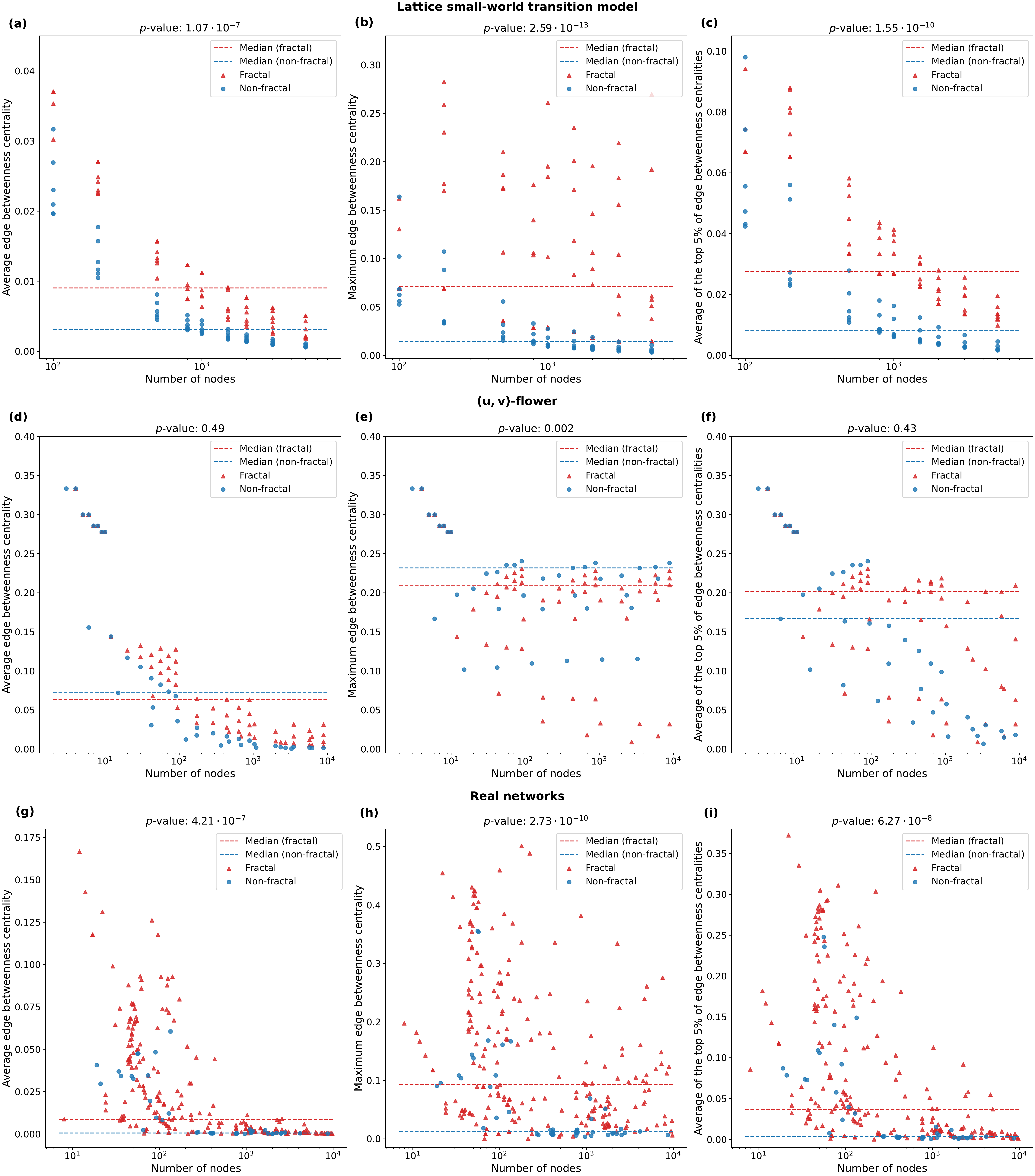}
    \caption{Measures concerning the edge betweenness centralities of \textbf{(a)}-\textbf{(c)} the Lattice small-world transition model, \textbf{(d)}-\textbf{(f)} the $(u,v)$-flower, \textbf{(g)}-\textbf{(i)} the real networks. \textbf{(a)},\textbf{(d)},\textbf{(g)} Average edge betweenness centrality; \textbf{(b)},\textbf{(e)},\textbf{(h)} maximum edge betweenness centrality; \textbf{(c)},\textbf{(f)},\textbf{(i)} average of the top 5\% of edge betweenness centralities plotted against the logarithm of the number of nodes. The $p$-values in the title of each subfigure correspond to the Mann-Whitney U rank test performed on the corresponding metric values to see the statistical significance of the difference in the distribution of fractal and non-fractal networks.}
    \label{fig:ebc}
\end{figure}

The Song-Havlin-Makse, the Hub attraction dynamical growth, and the Repulsion based fractal models generate networks for which the examined measures range from 0 to $0.6$, and they decrease as the number of nodes increases. A difference in fractal and non-fractal networks can be observed in the aforementioned three models: fractal networks tend to obtain larger values than non-fractal networks of the same size (see the supplementary material~\cite{supplementary}). %However, this difference disappears on the general scale, consequently, no universal conclusion can be drawn concerning the relation of fractality to edge betweenness centrality.
Similar observations can be made on the Lattice small-world transition model (see Figure~\ref{fig:ebc}~\textbf{(a)}-\textbf{(c)}). However, here the edge betweenness centralities are low in general for all parameter settings, regardless of fractality. Furthermore, we can observe that as the value of parameter $p$ grows, i.e., as the network becomes less fractal, the betweenness centrality of its edges decreases which contradicts the conjecture.

%Contrary to the previous models, this does not contradict completely the statement that fractal networks have smaller edge betweenness centralities than non-fractal networks.
Contrary to the previous models, for $(u,v)$-flowers, for any given network size, the maximal edge $BC$ is typically larger for non-fractal than for fractal networks. The same observation can be made for the average of the top 5\% betweenness centralities for small networks (fewer than 100 nodes), however, for larger networks, this property disappears, and the values are higher for fractal networks than for non-fractals. Moreover, the average of all betweenness centralities also shows that fractal networks have a higher average edge $BC$.
%Furthermore, all of these differences can only be observed for a given network size. As the $p$-values of the test for the significance of the difference of distributions show as well, on the general scale usually no distinction can be made between fractal and non-fractal networks based on measurements related to their edge betweenness centralities.
The aforementioned results are well illustrated in Figure~\ref{fig:ebc}~\textbf{(d)}-\textbf{(f)}.

%Figure~\ref{fig:ebc}~\textbf{(g)}-\textbf{(i)} suggests that no patterns can be observed concerning the edge betweenness centralities of fractal or non-fractal real networks since the two categories cannot be distinguished based on these edge betweenness centrality related measures. Moreover, no pattern can be recognized in any of the examined measures, even if we investigate the continuous measure of fractality, the $R$ coefficient (see the supplementary material~\cite{supplementary}).
Figure~\ref{fig:ebc}~\textbf{(g)}-\textbf{(i)} suggests that fractal real networks often have larger values concerning these edge betweenness centrality-related measures than non-fractals. However, there are also examples of non-fractal networks with high edge $BC$ measures, and when the values are low the fractal and non-fractal categories cannot really be distinguished. Moreover, similar observations can be made, even if we investigate the continuous measure of fractality, the $R$ coefficient (see the supplementary material~\cite{supplementary}).

To sum up, we can conclude that fractal networks can have edges with high betweenness centrality as well, furthermore, the related metric values seem to be higher on average for the fractal than non-fractal networks, which contradicts the suggested connection between edge betweenness and fractality.
%and the characteristics related to this metric seem to be independent of fractality.

\subsection*{Correlation of degree and betweenness centrality}
In one of the earliest works on fractal networks, Kitsak \textit{et al.}~\cite{kitsak2007betweenness} studied the betweenness centrality of fractal and non-fractal networks. The authors analyzed seven SHM model-generated networks and four relatively large real-world networks. They have found that there is a smaller correlation between the betweenness centrality and the degree of a node in fractal networks than in non-fractal networks. The authors argue that the Pearson correlation coefficient is not a suitable metric to characterize the difference between fractal and non-fractal networks because the average betweenness centrality for a given degree does not change much \cite{kitsak2007betweenness}. Hence, the authors measure the standard deviation of the betweenness centralities for a given degree and they compare it to that of the counterpart networks made by the configuration model. Due to the large computing complexity, here we apply a slightly different approach.

In this work, we measure the coefficient of variation (ratio of standard deviation to the mean) of the betweenness centralities for given degrees and then take the average along the degrees. A low mean coefficient of variation (CV) means that the correlation between the betweenness centrality and the degree is high, and similarly, a high CV means that the correlation is low. We also computed the Pearson correlation, and a weighted mean of the coefficient of variation, where similarly to Kitsak \textit{et al.}~\cite{kitsak2007betweenness}, we weighted by the degree distribution.
%We found that the different metrics for measuring the correlation between betweenness centrality and degree gives consistent results (for more details see the supplementary material~\cite{supplementary}). 

\subsubsection*{Results}

\begin{figure}[h!]
    \centering
    \includegraphics[width=.45\textwidth]{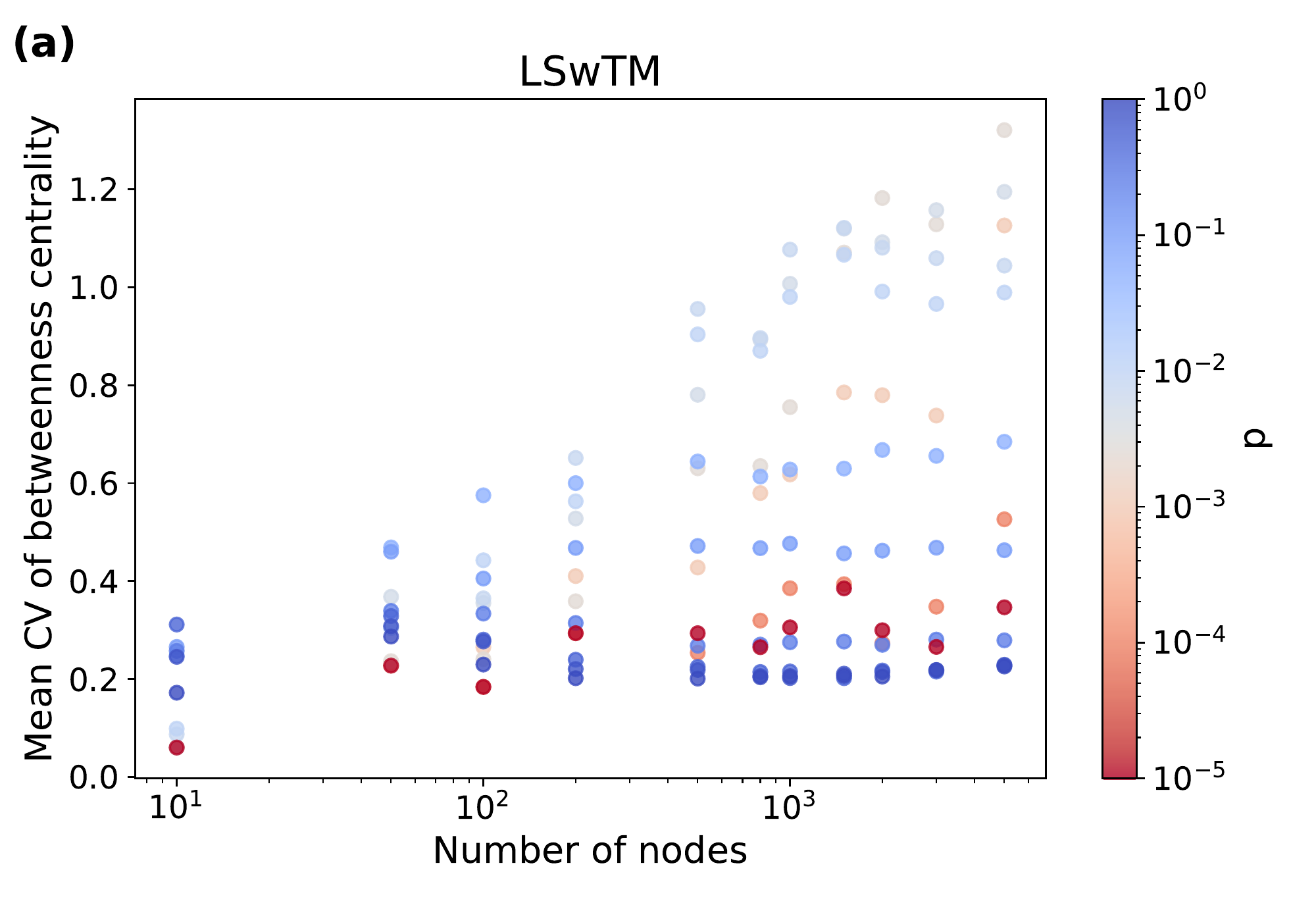}
    \includegraphics[width=.45\textwidth]{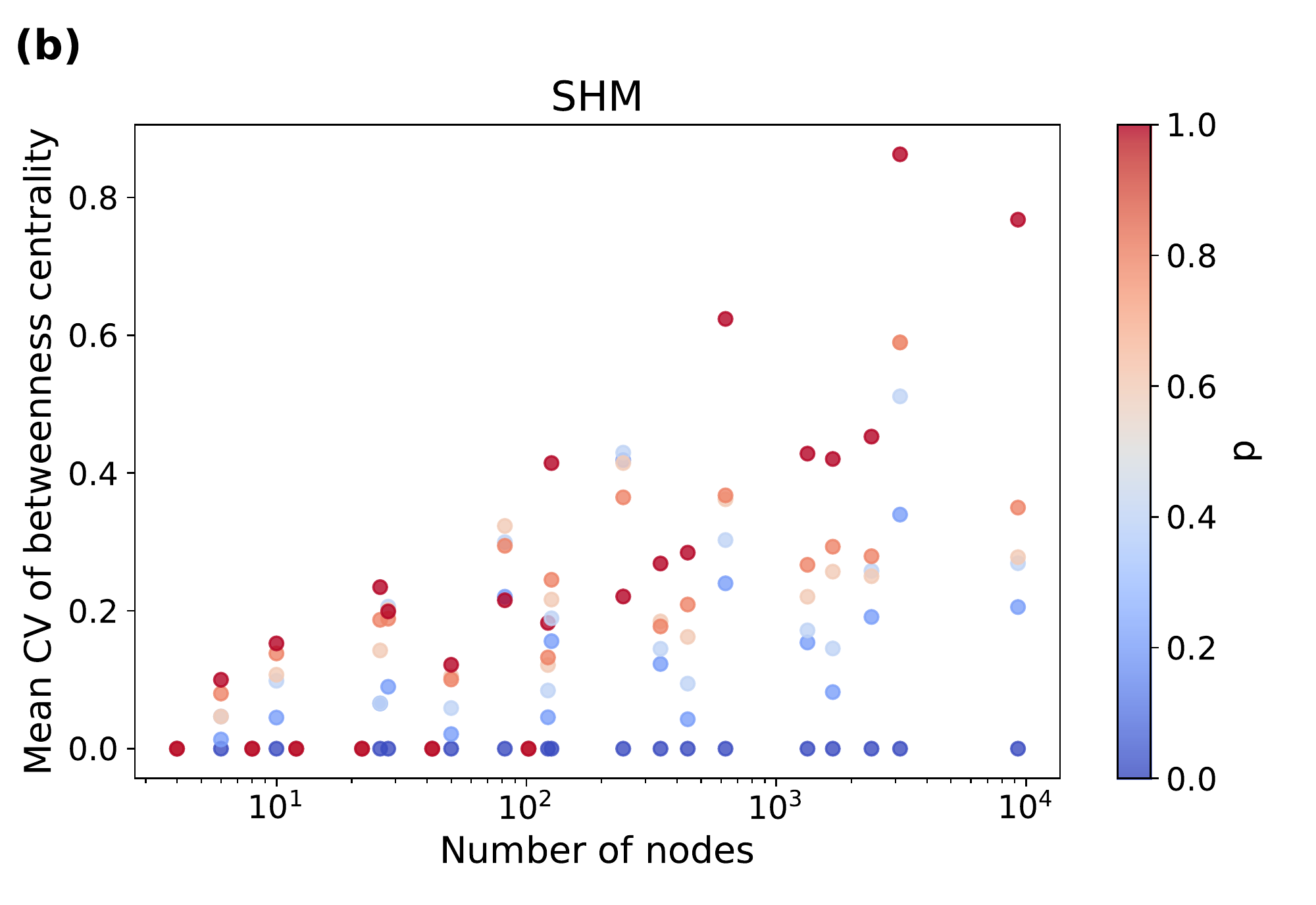}
    \includegraphics[width=.45\textwidth]{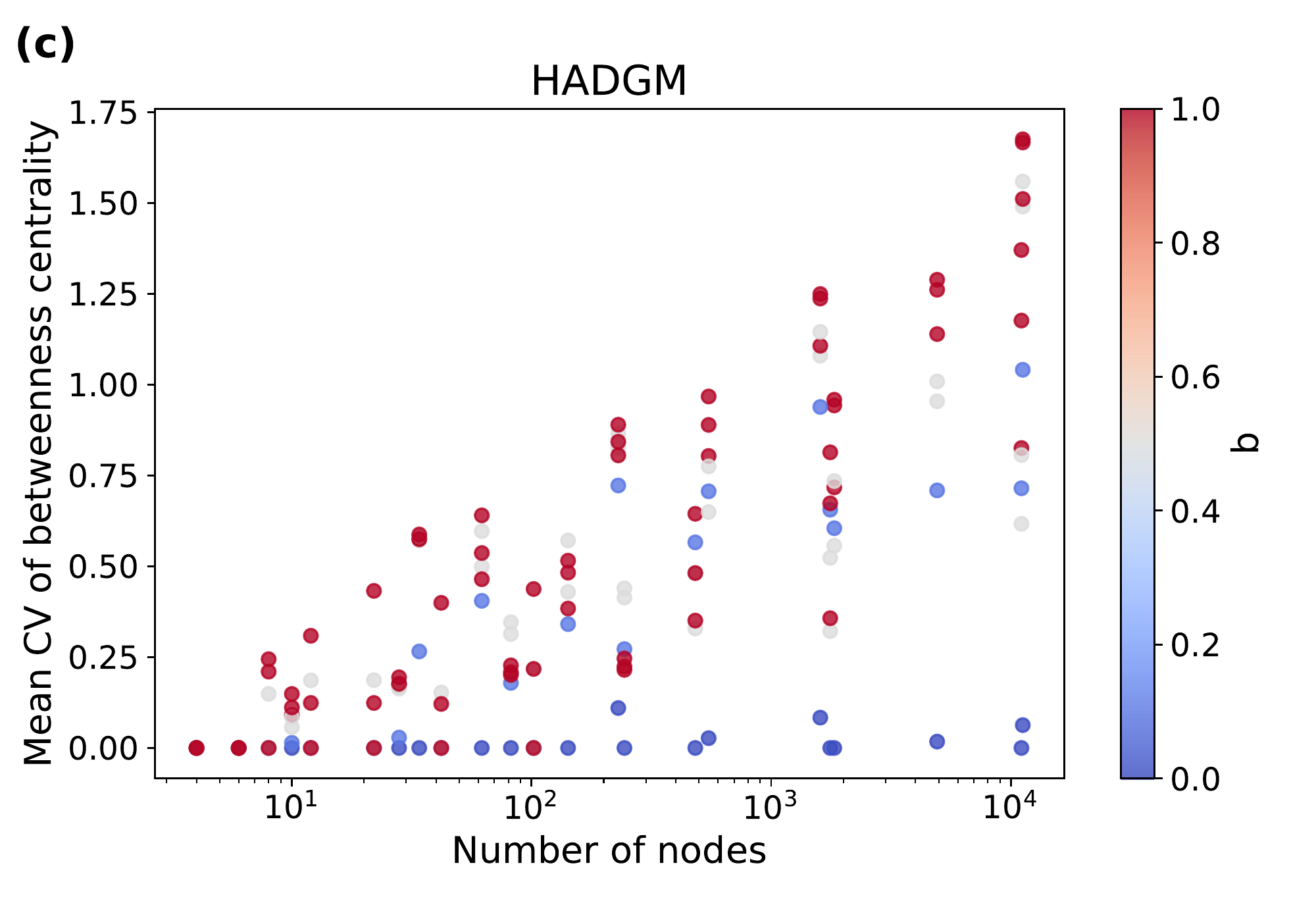}
    \includegraphics[width=0.45\textwidth]{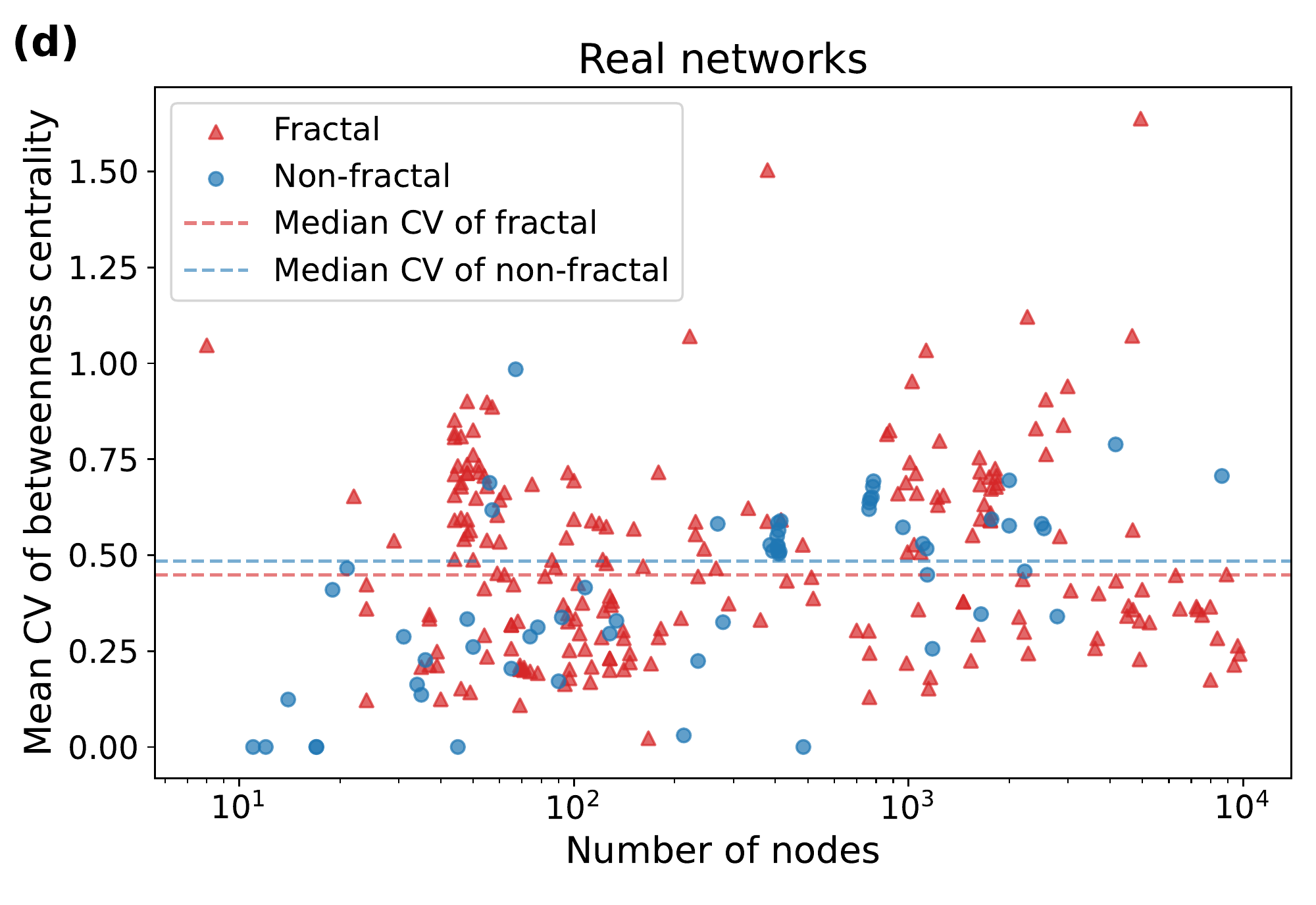}
        
    \caption{The mean coefficient of variation (CV) of the betweenness centrality per degree. The fractal networks are colored red and the non-fractal networks are blue. The difference in the distribution of the mean CV of the fractal and non-fractal real networks is not statistically significant (Mann-Whitney U rank test's $p$-value: 0.125).}
    \label{fig:btwc_cv}
\end{figure}

We found that the different metrics for measuring the correlation between betweenness centrality and degree gives consistent results. Here we discuss our findings concerning the mean CV in detail, for results about the other metrics see the supplementary material~\cite{supplementary}. 
Figure \ref{fig:btwc_cv} shows the mean CV of the betweenness centralities for three network models and the real-world networks. In the case of network models, for each parameter setting, we took the average of the results of 15 graphs.

%In the figures, the fractal networks are colored red and the non-fractal networks are blue. 
Figure \ref{fig:btwc_cv} suggests that the Song-Havlin-Makse model (Figure \ref{fig:btwc_cv} \textbf{(b)}) indeed supports the conjecture of Kitsak \textit{et al.} (for networks with at least 100 nodes), but the other network models and real-world networks do not seem to be in alignment with this conjecture. 

As Figure \ref{fig:btwc_cv} \textbf{(a)} illustrates, in the case of the LSwT model, the dispersion (CV) of betweenness centrality is low for both pure fractal and pure non-fractal networks. Surprisingly, those networks have the highest dispersion (i.e., lowest correlation) that possess a mixture of the fractal and non-fractal properties ($0<p<1$). 

In the case of the HADG model (Figure \ref{fig:btwc_cv} \textbf{(c)}) while it is true that the purely non-fractal network has lower variance (i.e., the higher correlation between degree and betweenness centrality) and that the purely fractal network has high dispersion. It is also apparent that when $b=0.1$ the model is still non-fractal, but the deviation of the betweenness centralities is as high as in the case of the fractal networks ($b=0.5$ and $b=1.0$). 

We also studied the $(u, v)$-flower (the results are included in the supplementary material~\cite{supplementary}), and we found that while on average the non-fractal networks ($u=1$) have lower CV values, it is also possible to generate fractal networks that have nearly zero variance regarding the betweenness centralities for given degrees. Hence, the $(u,v)$-flower model also does not really support the observation of Kitsak \textit{et al.}~\cite{kitsak2007betweenness}. 

Finally, Figure \ref{fig:btwc_cv} \textbf{(d)} demonstrates that in real networks the distribution of the coefficient of variation of the betweenness centrality in fractal and non-fractal networks does not differ.

\subsection*{Small-world property}\label{section:smallworld}
Another widely studied topic is the relationship between the small-world and fractal properties of networks. Csányi and Szendrői suggest that these two are conflicting properties of networks, however, they also mention that mixed property could also be possible in such a way that a network is microscopically small-world, but fractal on a macroscopic scale \cite{csanyi2004fractal_sw12}. Several other papers also reported that there is a connection between the lack of the small-world property and the emergence of fractality  \cite{zheng2014simple_sw2, uvflower_sw8, tian2008scaling_sw11, mokhlissi2020structural_sw13, zhang2008fractal_sw14}. On the other hand, plenty of models have been introduced, that exhibit transition from fractal to small-world networks \cite{ SHM_origins_sw9, li2017fractal_sw3, watanabe2015fractal_sw4, rozenfeld2010small_sw10, zhang2008transition_sw15}. Moreover, modifications of existing models have been proposed to demonstrate the simultaneous presence of the two properties in the generated networks \cite{SHM_origins_sw9, support_disassortativity_sw6, barriere2006fractality_sw5, kim2007fractality_sw7, ikeda2020fractality_sw1}.

The possible connection between the fractal and small-world property of networks has received a great deal of research interest, however, it has to be mentioned that the term \say{small-world network} is often used non-rigorously. By definition, a network can be considered small-world, if $l \sim \log N$
holds \cite{watts1998collective}, i.e., when the average distance in the network ($l$) grows proportionally to the logarithm of the number of nodes ($N$). However, this definition can only be applied to networks evolving with time or where different states can be compared, so typically to network models. Consequently, networks having a relatively small diameter or average path length compared to their size are also often referred to as small-world networks~\cite{porter2012small}.

Here, we rather distinguish between the two approaches and examine the relation of fractality to the small-world property using two slightly different approaches. First, we consider the length of the diameter and average path length for all networks to see whether fractal networks have larger distances. In the second approach, we consider growing network models to study whether the original concept of small-world property and the fractal property can simultaneously be present in a network or whether these are exclusive characteristics.

\subsubsection*{Results with normalized average path length and diameter}
For the first approach, to be able to compare the distances of networks of different sizes, both the average path length and the diameter are normalized by the logarithm of the number of nodes~\cite{nagy2022network}. 

\begin{figure}[h!]
    \centering
    \includegraphics[width=0.95\textwidth]{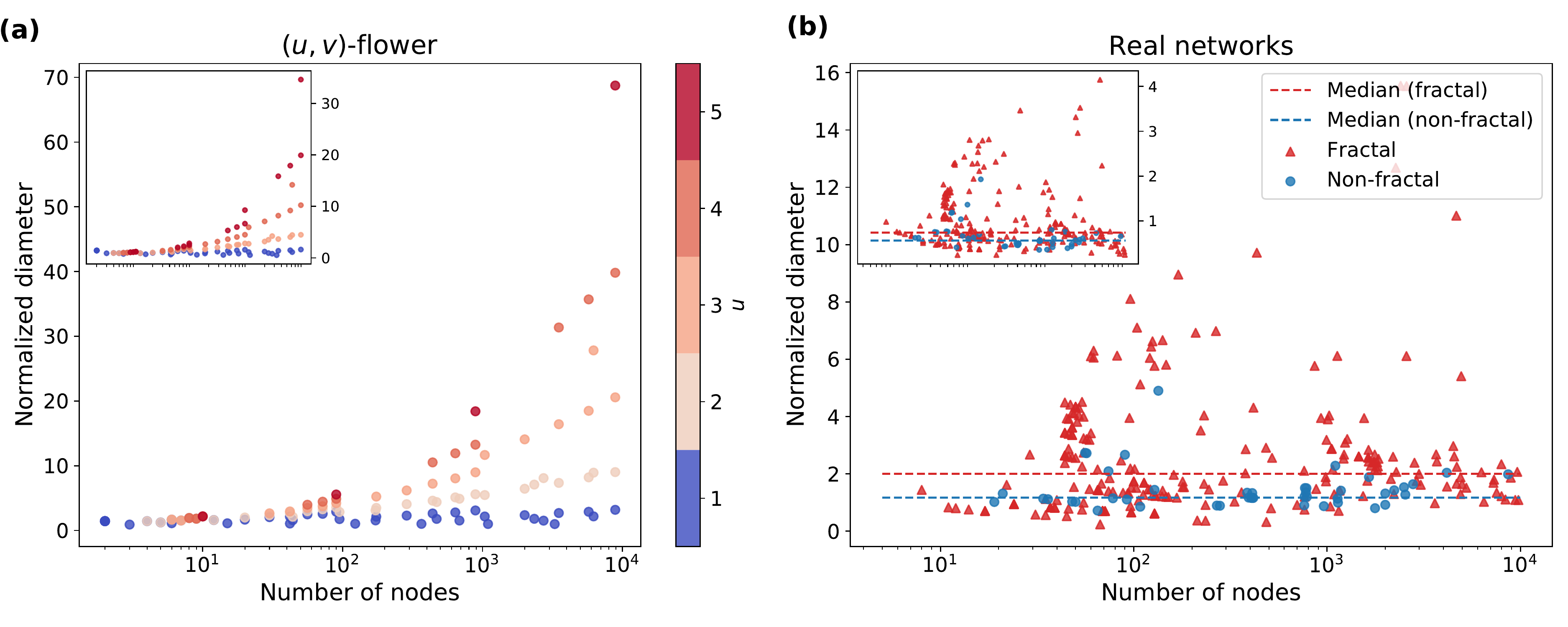}
    \caption{Normalized diameter (i.e. diameter divided by the logarithm of the size) plotted against the size on a semi-log scale for the cases of the \textbf{(a)} $(u,v)$-flower, \textbf{(b)} real networks. On subplot \textbf{(a)} the coloring is based on the $u$ parameter of the model. The insets are showing the semi-log plots of the normalized average path length plotted against the size of the network. The difference between the distribution of normalized diameter and the normalized average path length of the fractal and non-fractal real networks are statistically significant (Mann-Whitney U rank test's $p$-value: $7.66\cdot10^{-6}$ and $1.19\cdot10^{-5}$). The results for the other models, as well as an $R$-value-based illustration of the real networks, can be found in the supplementary material~\cite{supplementary}.}
    \label{fig:avgpath_diam}
\end{figure}

In the case of the $(u,v)$-flower and the Lattice small-world transition model, apart from networks with very few nodes, the different cases determined by the main parameter ($u$ for the $(u,v)$-flower and $p$ for the LSwTM) clearly separate, and the distances are growing as the networks become fractal. This phenomenon is well illustrated by Figure~\ref{fig:avgpath_diam}~\textbf{(a)}, which shows the change in the normalized diameter and average path length for the $(u,v)$-flower. 

In the case of the Song-Havlin-Makse and the HADG model, this clear separation can only be observed for fixed values of the $(n,m)$ parameter pair. Namely, the parameter, which influences fractality, also affects the distances similarly to LSwTM and $(u,v)$-flower. %However, fractal and non-fractal networks cannot be clearly separated globally. 
%Increasing the value of this parameter, hence making the generated network more fractal, results in the increase of the distances. 

Since the Repulsion based fractal model always generates fractal networks, the previous comparing-based examination cannot be applied to this model, however, it can be said that the average path length and the diameter of this model are similar to those of the fractal networks generated by the SHM and HADG models. 
% except in a few cases, the distances are generally small independently of the size and the parameter setting. The normalized average path length is usually less than 3, and the normalized diameter typically does not exceed 7.

As Figure~\ref{fig:avgpath_diam}~\textbf{(b)} illustrates, similar observations can be made on the real networks, too. The distances are mostly quite small both in non-fractal and fractal networks and for small values the two classes can hardly be separated based on the normalized diameter and average path length. 
%these measurements. 
However, we can observe that non-fractal networks do not tend to have large distances, and based on our dataset, a cut-off point at 2.5 for the normalized average path length, and at 5 for the normalized diameter can be created. Above these values, there seem to be only fractal networks. However, we have to emphasize that there are numerous fractal networks below these cut-off points as well, which means that fractality does not originate just from large distances.

\subsubsection*{Results with growing network models}
\begin{figure}[h!]
    \centering
    \includegraphics[width=0.95\textwidth]{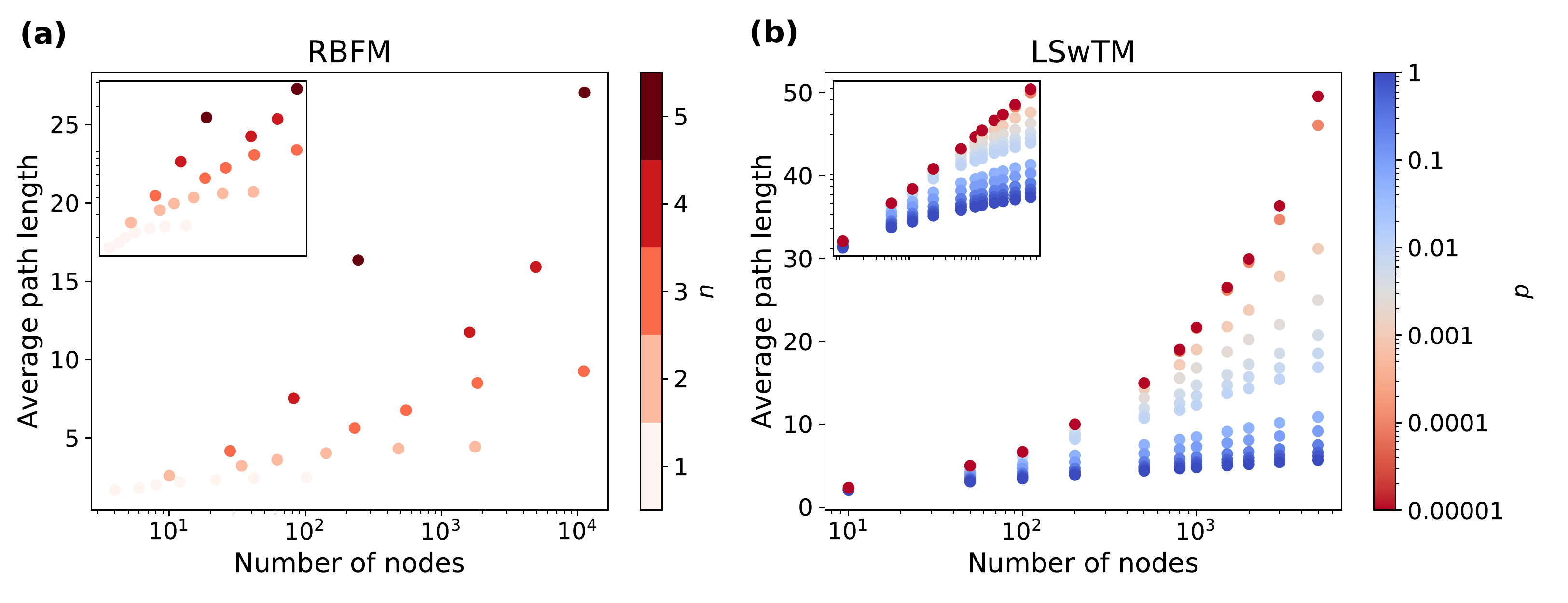}
    \caption{Average path length as a function of the logarithm of the size for some cases of the \textbf{(a)} Repulsion-based fractal model, \textbf{(b)} Lattice small-world transition model. On subplot \textbf{(a)} the networks are generated by setting parameter $Y$ to 0, and the coloring is based on parameter $n$. On subplot \textbf{(b)} the coloring is based on parameter $p$. The insets are showing the log-log plots of the same cases.}
    \label{fig:smallworld}
\end{figure}

%Here the relation of the average path length to the size is examined, in the case of the network models.
All the investigated network models undergo a transition from the small-world to the non-small-world property driven by the main parameter that also drives the fractality of the network: as fractality weakens, small-world property arises.
%Considering all of the networks generated by a given model, and comparing the cases determined by the main parameter, which influences fractality, a transition from the small-world to the non-small-world property can be observed.
This transition is not sharp, and there are intermediate states, where both properties are significant. The small-world transition can also be observed in the Repulsion based fractal model, which generates only fractal networks. Figure~\ref{fig:smallworld} illustrates some cases of the Repulsion based fractal model and the Lattice small-world transition model. It can be seen that there are a few states for both models, where the small-world property is met, while the networks are also fractal.

Except for the Lattice small-world transition model, the relation of the average path length to the size can also be examined considering the iterations, through which the network evolves. That is we fix all the parameters, except the iteration number ($n$), and see if the average path length grows proportionally to the logarithm of the size or power law holds instead. It can be said that in this approach, for the cases where the resulting networks are rather fractal than non-fractal, indeed power law holds, while in the non-fractal cases logarithmic relation can be observed. The only exception is the Repulsion based fractal model because it has a few (fractal) cases, where the connection is rather logarithmic, for which an illustration can be found in the supplementary material~\cite{supplementary}. 
%Consequently, it seems that small-world property does not preclude a network to be fractal as well.

Overall, if we take into consideration both of the approaches concerning the small-world property we can say that there is a significant relation between fractality and non-small-world property, however, they do not necessarily exclude each other since there are examples for networks, which are fractal and small-world at the same time.

\section*{A machine learning approach}\label{section:ml}
In the previous sections, we investigated the network characteristics that have been associated with fractality one by one. Here, we address the problem as a binary classification task and distinguish between fractal and non-fractal networks based on a few selected network characteristics. The benefit of this approach is that we can investigate numerous network metrics at the same time, identify the most important features and also recognize how the combination of metrics affects the fractal scaling of networks.

%propose a different approach to gain a better understanding of the underlying mechanisms behind fractality. Namely,  and try to. We intend to, with the help of machine learning algorithms, identify the most important features.  %In this way, we could conclude which properties have the most significant relationship with the fractal scaling of networks.

Here, we use three decision tree-based classification algorithms to distinguish between fractal and non-fractal networks:  simple decision tree, random forest, and XGBoost. We select the explanatory variables to get a collection of characteristics, which represents the structure of the networks well, but they are not too correlated. Moreover, we aim to make these metrics as independent of the network size as possible, hence where it is reasonable, normalization is also performed. We extend the set of metrics that we used in our earlier studies \cite{nagy2022network, nagy2019structural} with features that have been associated with fractality. The list of our explanatory variables together with their description can be found in Table~\ref{table:metrics}.

\begin{table}[h!]
\centering
\scalebox{1}{
\begin{tabular}{l|l}
\multicolumn{1}{c|}{\multirow{2}{*}{\textbf{Name}}} & \multicolumn{1}{c}{\multirow{2}{*}{\textbf{Description}}} \\
\multicolumn{1}{c|}{} & \multicolumn{1}{c}{} \\ \hline
avg\_deg &  Average degree\\ 
assortativity & Assortativity coefficient \\ 
avg\_clust & Average of the local clustering coefficients\\ 
max\_eigen & Maximum of the eigenvector centralities \\ 
skew\_deg\_dist & Skewness of the degree distribution \\ 
diam\_logsize & Diameter, divided by the logarithm of the number of nodes \\ 
max\_deg\_n & Maximum degree, divided by the number of nodes minus 1 \\
hub\_connectivity & Hub connectivity score \\
ebc\_avg\_95 & \begin{tabular}[l]{@{}l@{}}Average of the edge betweenness centralities\\ over the top 5\% of edges \end{tabular}
\end{tabular}}
\caption{Name and description of the chosen explanatory variables, for the classification task.}
\label{table:metrics}
\end{table}

We consider three datasets to perform the task on, one consisting of the model-generated networks, one of the real networks, and one which combines the two sets, thus including all examined networks. Moreover, we drop the small networks from all datasets, i.e. the ones whose number of nodes is less than 100, because in most of these cases fractality can hardly be defined, as was also mentioned earlier. We use 2/3 of the datasets for training and the remaining 1/3 for testing to avoid overfitting. Two evaluation metrics are used to measure the performance of the algorithms, accuracy and the Area Under the ROC Curve (AUC). 
%To read about the definition and usage of these measures, as well as the basics of data science, we suggest the following introductory book: \cite{tan2016introduction}. 
The hyperparameter optimization for the algorithms is carried out based on the latter one because, in the case of an unbalanced class distribution, the accuracy score can often be misleading. For the data preparation, training, and evaluation of the algorithms, we use the  \textit{scikit-learn} \cite{scikit-learn} and \textit{XGBoost} \cite{xgboost} Python packages.

To identify the most important variables, we calculate the \textit{permutation importance} score of the features. This score shows how much the performance of the model decreases if the values of a given attribute are randomly permuted. %For details concerning the determination of the scores, please refer to the documentation of the \textit{permutation importance} function %\footnote{\url{https://scikit-learn.org/stable/modules/permutation_importance.html}}
%of the \textit{scikit-learn} package. 

\subsubsection*{Results}
The performance of the models measured on the test sets is summarized in Table~\ref{table:ml_results}. It can be said that all of the algorithms can solve the problem with high accuracy and AUC score, thus we can conclude that fractal and non-fractal networks indeed differ in the considered network characteristics.

\begin{table}[h]
\scalebox{1}{
\begin{tabular}{l|ccc|ccc|ccc}
\multicolumn{1}{c|}{\multirow{2}{*}{\textbf{}}} & \multicolumn{3}{c|}{\textbf{Decision tree}} & \multicolumn{3}{c|}{\textbf{Random forest}} & \multicolumn{3}{c}{\textbf{XGBoost}} \\ \cline{2-10} 
\multicolumn{1}{c|}{} & \textbf{model} & \textbf{real} & \textbf{all} & \textbf{model} & \textbf{real} & \textbf{all} & \textbf{model} & \textbf{real} & \textbf{all} \\ \hline
\textbf{Accuracy} & 0.97 & 0.86 & 0.90 & 0.95 & 0.91 & 0.95 & 0.95 & 0.86 & 0.95 \\
\textbf{AUC} & 0.98 & 0.90 & 0.97 & 1.00 & 0.95 & 0.98 & 0.99 & 0.82 & 0.98
\end{tabular}}
\centering
\caption{Accuracy and AUC scores of the different algorithms on the fractal/non-fractal classification task, on the three  datasets.}
\label{table:ml_results}
\end{table}

\begin{table}[h]
\scalebox{0.9}{
\begin{tabular}{ll|ccc}
 & \multicolumn{1}{c|}{\textbf{}} & \textbf{First feature} & \textbf{Second feature} & \textbf{Third feature} \\ \hline
\multicolumn{1}{l|}{\multirow{3}{*}{\textbf{model}}} & \textbf{DT} & diam\_logsize (0.285) & hub\_connectivity (0.198) & assortativity (0.109) \\
\multicolumn{1}{l|}{} & \textbf{RF} & hub\_connectivity (0.136) & diam\_logsize (0.071) & avg\_clust (0.020) \\
\multicolumn{1}{l|}{} & \textbf{XGB} & diam\_logsize (0.250) & hub\_connectivity (0.159) & assortativity (0.071) \\ \hline
\multicolumn{1}{l|}{\multirow{3}{*}{\textbf{real}}} & \textbf{DT} & avg\_deg (0.107) & hub\_connectivity (0.104) & diam\_logsize (0.067) \\
\multicolumn{1}{l|}{} & \textbf{RF} & avg\_clust (0.040) & assortativity (0.038) & max\_deg\_n (0.029) \\
\multicolumn{1}{l|}{} & \textbf{XGB} & assortativity (0.112) & hub\_connectivity (0.044) & diam\_logsize (0.022) \\ \hline
\multicolumn{1}{l|}{\multirow{3}{*}{\textbf{all}}} & \textbf{DT} & diam\_logsize (0.290) & assortativity (0.118) & hub\_connectivity (0.098) \\
\multicolumn{1}{l|}{} & \textbf{RF} & hub\_connectivity (0.086) & diam\_logsize (0.082) & avg\_clust (0.058) \\
\multicolumn{1}{l|}{} & \textbf{XGB} & diam\_logsize (0.272) & avg\_deg (0.089) & hub\_connectivity (0.088)
\end{tabular}}
\centering
\caption{The most important features for every algorithm and dataset. For all the cases the three features with the highest \textit{permutation importance} score are listed, showing the exact scores as well.}
\label{table:importance}
\end{table}

Table~\ref{table:importance} shows the three most important features with the corresponding \textit{permutation importance} scores for every algorithm and dataset. We can observe that for model-generated networks, the normalized diameter and the hub connectivity score both have significant importance for all algorithms. In addition, the assortativity coefficient also seems to be important for most of the methods.
In the case of real networks, the set of important features varies for the different machine learning (ML) models. The assortativity coefficient, the hub connectivity score, and the normalized diameter are among the most important attributes for two of the three algorithms, but the average clustering coefficient and the average or maximum degree can also be considered important features for some ML models. 
In the combined dataset, the normalized diameter and the hub connectivity score turned out to be the most important characteristics of all algorithms. The assortativity coefficient, average clustering coefficient, and average degree also seem to have notable importance for some of the methods. 

\begin{figure}[h!]
    \centering
    \includegraphics[width=0.9\textwidth]{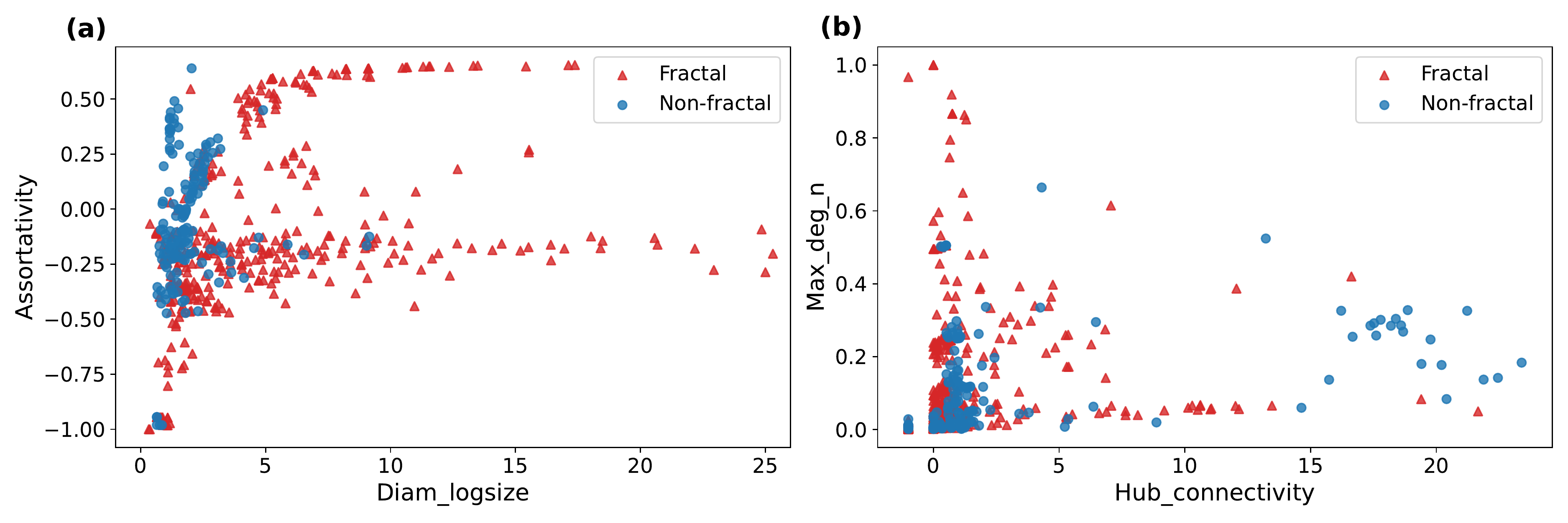}
    \caption{Scatter plots of the dataset consisting of both real and model generated networks, with respect to two selected important features. \textbf{(a)} Assortativity coefficient plotted against the normalized diameter. \textbf{(b)} Normalized maximum degree plotted against the hub connectivity score.}
    \label{fig:ML}
\end{figure}

Figure \ref{fig:ML} shows two scatterplots of the combined dataset with respect to different network characteristics. It can be seen that while a large (normalized) diameter is a characteristic of only the fractal networks, an additional feature is still not enough to clearly separate fractal and non-fractal networks when the diameter is small. Similarly, most of the investigated fractal networks possess a small hub connectivity score, but there are a significant number of them with large $HCS$, and considering the normalized maximum degree as well, we still cannot separate the two classes clearly.

From the results detailed above, we can conclude that the magnitude of distances in a network is certainly connected to fractality. It may not be the distances between hubs, which influence fractality, but average distances generally. However, the connectivity of hubs, as well as assortativity indeed seem to have a distinguishing ability. Although alone they are not enough to separate fractal and non-fractal networks, together with other properties, they could contribute to the distinction. It seems that, although a single network characteristic does not clearly determine the fractal property, the combination of several metrics can achieve excellent distinguishing power.

\section*{Summary}\label{chapter:4}
In this work, we investigated which characteristics could cause the emergence of fractal scaling in complex networks. Our analyses relied on a large dataset of both real-world and model-generated networks, in order to prevent making conclusions based on coincidences. Our most important findings are summarized in Table~\ref{table:summary}.

\begin{table}[h]
\scalebox{0.65}{
\begin{tabular}{ll|cccccc}
\multicolumn{2}{l|}{} & \textbf{SHM} & \textbf{HADGM} & \textbf{$(u,v)$-flower} & \textbf{RBFM} & \textbf{LSwTM} & \textbf{Real networks} \\ \hline
\multicolumn{2}{l|}{\textbf{Disassortativity}} & \makecell{frac.$\Longrightarrow$disass. \\ disass.$\centernot\Longrightarrow$frac.} & \makecell{frac.$\Longrightarrow$disass. \\ disass.$\centernot\Longrightarrow$frac.} & \cmark & \cmark & \xmark &  \cmark \\
\multicolumn{2}{l|}{\textbf{Hub repulsion}} & \cmark & \cmark & \cmark & $\bigcirc$ & \cmark &  \cmark \\ \hline
\multirow{3}{*}{\textbf{\begin{tabular}[c]{@{}l@{}}Long range \\ anticorrelation\end{tabular}}} & \textbf{\begin{tabular}[c]{@{}l@{}}Fluctuation\\ analysis\end{tabular}} & \cmark & $\bigcirc$ & \cmark & \xmark & \xmark & $\bigcirc$ \\
 & \textbf{Hub distances} & $\bigcirc$ & \cmark & \makecell{\cmark ($n>3$)} & \cmark & -- & $\bigcirc$ \\
 & \textbf{\begin{tabular}[c]{@{}l@{}}Neighbor\\ connectivity\end{tabular}} & $\bigcirc$ & $\bigcirc$ & $\bigcirc$ & $\bigcirc$ & \xmark & $\bigcirc$ \\ \hline
\multirow{3}{*}{\textbf{Small edge BC}} & \textbf{Mean EBC} & \xmark & \xmark & \xmark & \xmark & \xmark & \xmark \\
 & \textbf{Max EBC} & \xmark & \xmark & \cmark & \xmark & \xmark & \xmark\\
 & \textbf{Mean of top 5\%} & \xmark & \xmark & \begin{tabular}[c]{@{}c@{}}\cmark ($N\leq100$)\\ \xmark ($N>100$)\end{tabular} & \xmark & \xmark & \xmark \\ \hline
\multicolumn{2}{l|}{\textbf{\begin{tabular}[c]{@{}l@{}}Weak correlation between \\ degree and node BC\end{tabular}}} & \cmark ($N>100$) & $\bigcirc$ & $\bigcirc$ & $\bigcirc$ & $\bigcirc$ & $\bigcirc$ \\ \hline
\multirow{2}{*}{\textbf{\begin{tabular}[c]{@{}l@{}}Non-small-world \\ property\end{tabular}}} & \textbf{\begin{tabular}[c]{@{}l@{}}Average path length\\ and diameter\end{tabular}} & \cmark & \cmark & \cmark & $\bigcirc$ & \cmark & \begin{tabular}[c]{@{}c@{}}large dist. $\Longrightarrow$ frac.\\ small dist. $\centernot\Longrightarrow$ non-frac.\end{tabular} \\
 & \textbf{\begin{tabular}[c]{@{}l@{}}Growing network\\ models\end{tabular}} & $\bigcirc$ & $\bigcirc$ & $\bigcirc$ & $\bigcirc$ & $\bigcirc$ & --
\end{tabular}}
\centering
\caption{The relation of network models and real networks to the examined conjectures concerning fractal characteristics. The check mark (\cmark) denotes that the model/real networks support the statement (for real networks it means that the distribution of the metric is statistically significantly different in fractal and non-fractal networks), the cross mark (\xmark) means that they contradict it, i.e. they behave in the opposite way, the circle ($\bigcirc$) refers to mixed results and when no clear conclusion can be drawn, moreover, the hyphen (--) denotes when the analysis could not be carried out due to the nature of the model. Finally, $N$ denotes the number of nodes and $n$ denotes the number of iterations.}
\label{table:summary}
\end{table}
% \footnotetext{In the case of real networks, check mark means, that the distribution of the metric is statistically significantly different.}

Concerning the disassortativity of fractal networks, we have found that although most of the considered mathematical models suggest that fractality correlates with disassortativity, there is also one model, the Lattice small-world transition model, which completely contradicts the statement. Consequently, we can conclude that although disassortativity is common amongst fractal networks, just based on the disassortativity, we cannot clearly tell whether a network is fractal or not,  which is suggested by real networks as well. We conclude that disassortativity cannot be considered the reason behind fractality,  %Consequently, we can conclude that there is no clear correlation between assortative mixing and fractality, which is suggested by the real networks as well. 
Somewhat similar observations can be made in the case where hub repulsion was measured directly. All of the considered network models show that in fractal networks hubs are less connected than in non-fractal networks (smaller hub connectivity score). The real networks also suggest that a large hub connectivity score (hub attraction) is a property of non-fractal networks, but counterexamples on both sides make hub repulsion a non-universal characteristic of fractal networks.
%Although all of the considered network models show that in fractal networks hubs are less connected than in non-fractal networks (smaller hub connectivity score), however, real-world fractal networks cannot be distinguished efficiently from non-fractals based on their hub repulsion characteristic.

The possible connection of long-range anticorrelation to fractality was reviewed using three different methods. 
%Just as in the case of neighbor-level anticorrelations, 
Here we could not find a clear connection between the correlation of node degrees and  fractal scaling. Although for all three methods, we could find examples of both model-generated and real networks, which support the suggestion of anticorrelation in fractal networks, even on the long-range scale, there are numerous counterexamples as well.

The suggestion of the connection of edge betweenness centrality with fractality was also reviewed. We examined whether fractal networks can possess edges with large betweenness centrality. We have come to the conclusion that fractal networks show no tendency to have edges mostly with small betweenness centrality. Almost all of the examined network models show the opposite of the statement, while real networks suggest that small edge betweenness centrality does not depend on fractality. 

In addition to the connection of fractality and edge betweenness centrality, a suggestion regarding node betweenness centrality was also revised. Namely, we revisited the conjecture that the correlation between degree and node BC is weaker in fractal networks than in non-fractals. We have found that the Song-Havlin-Makse model supports this statement, but all the other mathematical models and the real networks rather contradict it.
%but all the other mathematical models and the real networks show that the property of having a weak correlation between degree and BC is independent of fractality.

We investigated thoroughly the suggested conflicting relation of fractality and the small-world property. We have found that those network models which are able to generate both fractal and non-fractal instances support the observation that the distances (average path length and diameter) are larger in fractal networks. In the case of real networks, we have also found that large distances are present only in fractal networks, however, small distances do not imply non-fractality. Moreover, a transition from \say{small-worldness} to large distances can be observed in the Repulsion based fractal model as well, which generates only fractal networks. When we examined the small-world property on growing network models we experienced that the vast majority of the considered model-generated networks support the conflicting relation of fractality and small-world property.

Finally, we introduced a novel approach to analyze the origin of fractal networks. We formulated a binary classification problem with the goal to distinguish fractal and non-fractal networks based on other network properties. We solved the problem with state-of-the-art machine learning algorithms and identified the characteristics with high distinguishing ability. The results suggest that although a single characteristic is not enough, a combination of several metrics can distinguish between fractal and non-fractal networks efficiently. The normalized average distance is possibly one of the most essential properties in recognizing fractal scaling, moreover, hub connectivity and assortativity can also contribute to the characterization of fractal networks together with other properties.

An important direction of further studies could be to directly examine the possible connection of the proposed joint properties to fractality. For these analyses, the extension of the dataset with additional models and real networks may be necessary. Furthermore, other network characteristics should be involved in such studies, which have not been considered in previous works. Different approaches could also be used to distinguish fractal and non-fractal networks, such as network embedding techniques. If the networks could be embedded into a vector space, where the two classes are well-separated, then the properties of this space could reveal what the difference lies in.

%\nocite{oreg,schn,pond,smith,marg,hunn,advi,koha,mouse}

%%%%%%%%%%%%%%%%%%%%%%%%%%%%%%%%%%%%%%%%%%%%%%
%%                                          %%
%% Backmatter begins here                   %%
%%                                          %%
%%%%%%%%%%%%%%%%%%%%%%%%%%%%%%%%%%%%%%%%%%%%%%

\begin{backmatter}

\section*{Competing interests}
\noindent The authors declare that they have no known competing financial interests.

\section*{Author's contributions}

    RM conceived the study. EZP reviewed the literature. EZP and MN implemented and carried out the analyses. EZP prepared the original draft. MN and RM reviewed and edited the manuscript. All authors read and approved the final manuscript. 

\section*{Acknowledgment}
\noindent This work has been partially supported by the National Research, Development, and Innovation Office (NKFIH, Project K142169) and by the "Fractal geometry and applications" Research Group (NKFIH, Project KKP144059).

\section*{Data availability}
\noindent The networks and the data analyzed in this study are publicly available at the following GitHub repository \cite{supplementary}: \url{https://github.com/marcessz/fractal-networks} 
%%  After submission of the .TEX file,                     %%
%%  you will be prompted to submit your .BBL file.         %%

% if your bibliography is in bibtex format, use those commands:
\bibliographystyle{bmc-mathphys} % Style BST file (bmc-mathphys, vancouver, spbasic).
\bibliography{bmc_article}      % Bibliography file (usually '*.bib' )
% for author-year bibliography (bmc-mathphys or spbasic)
% a) write to bib file (bmc-mathphys only)
% @settings{label, options="nameyear"}

% or include bibliography directly:
% \begin{thebibliography}
% \bibitem{b1}
% \end{thebibliography}

%%%%%%%%%%%%%%%%%%%%%%%%%%%%%%%%%%%
%%                               %%
%% Figures                       %%
%%                               %%
%% NB: this is for captions and  %%
%% Titles. All graphics must be  %%
%% submitted separately and NOT  %%
%% included in the Tex document  %%
%%                               %%
%%%%%%%%%%%%%%%%%%%%%%%%%%%%%%%%%%%

%%
%% Do not use \listoffigures as most will included as separate files

%%%%%%%%%%%%%%%%%%%%%%%%%%%%%%%%%%%
%%                               %%
%% Tables                        %%
%%                               %%
%%%%%%%%%%%%%%%%%%%%%%%%%%%%%%%%%%%

%% Use of \listoftables is discouraged.
%%

%%%%%%%%%%%%%%%%%%%%%%%%%%%%%%%%%%%
%%                               %%
%% Additional Files              %%
%%                               %%
%%%%%%%%%%%%%%%%%%%%%%%%%%%%%%%%%%%

\end{backmatter}
\end{document}